\documentclass[aps,prd,nofootinbib,superscriptaddress]{revtex4-2}

\usepackage{amssymb}

\usepackage{mathrsfs}         
\usepackage{epsfig}
\usepackage{graphicx}
\usepackage{color}

\usepackage{amssymb}
\usepackage{fixltx2e}
\usepackage{bm}
\usepackage{amsmath,amsfonts,bm,dsfont}
\usepackage{verbatim}
\usepackage{mathrsfs}  %
\usepackage{definitions}
\usepackage{color} \definecolor{darkgreen}{rgb}{0,.5,0}
\usepackage{color}

\newcommand{\rv}[1]{{#1}
}

\def\rA{{\rm A}}
\def\rR{{\rm R}}
\def\rK{{\rm K}}
\def\rM{{\rm M}}
\def\rT{{\rm T}}

\begin{document}
\title{Wigner-Weyl calculus in Keldysh technique}
\author{C. Banerjee}
\email{banerjee.chitradip@gmail.com}
\affiliation{Ariel University, Ariel 40700, Israel}
\author{I.V. Fialkovsky}\email{fialkovsky.i@ufabc.edu.br}
\affiliation{CMCC-Universidade  Federal  do  ABC, 09210-170  Santo  Andre,  Brazil}
\author{M. Lewkowicz}\email{lewkow@ariel.ac.il}
\affiliation{Ariel University, Ariel 40700, Israel}
\author{C.X. Zhang}\email{zhang12345s@sina.com}
\author{M.A. Zubkov}
\email{mikhailzu@ariel.ac.il}
\affiliation{Ariel University, Ariel 40700, Israel}
\date{\today}
\begin{abstract}
We discuss the non-equilibrium dynamics of condensed matter/quantum field systems in the framework of Keldysh technique. In order to deal with the inhomogeneous systems we use the Wigner-Weyl formalism. Unification of the mentioned two approaches is demonstrated on the example of Hall conductivity. We express Hall conductivity through the Wigner transformed two-point Green's functions. We demonstrate how this expression is reduced to the topological number in thermal equilibrium at zero temperature. At the same time both at finite temperature and out of equilibrium the topological invariance is lost. Moreover, Hall conductivity becomes sensitive to interaction corrections.
\end{abstract}

\maketitle

\section{introduction}

Keldysh formalism \cite{Keldysh64} allows to investigate non-equilibrium phenomena both in condensed matter physics and in high energy physics \cite{Bonitz00,BS03,BF06} (see also \cite{KB62,Baym62,Schwinger61}). This formalism is actually the complete formulation of quantum field theory (QFT). The conventional formalisms of finite temperature equilibrium quantum statistical physics \cite{Matsubara55,BdD59,Gaudin60,AGD63}  and conventional real time QFT are used more often in condensed matter physics and in elementary particle physics, but they only represent the limiting particular cases of the more general Keldysh QFT. At the same time, this formalism contains the same basic ingredients as conventional QFT. In particular, the functional integral formalism \cite{Kamenev} may be used here, as well as the operator formalism \cite{Langreth76,Danielewicz84,CSHY85,RS86,Berges04,HJ98,Rammer07}. The essential difference from the conventional QFT is the appearance of the integration over the so-called Keldysh contour in the complex plane of time, $t$, instead of the integration over the real axis in this plane (the real time QFT) or imaginary axis (Matsubara technique of finite temperature equilibrium statistical mechanics).  It appears that unlike the conventional QFT the continuum functional integral formulation of Keldysh QFT faces certain difficulties related to the continuity of functions at the turning points of Keldysh contour. Therefore, operator approach to non-equilibrium diagram technique \cite{LP,Mahan} is still more popular.  Notice,  that in lattice regularization the precise and clear functional integral formulation is back \cite{Kamenev}.

Within Keldysh technique the perturbation theory has been developed similar to that of the conventional QFT \cite{Keldysh64,Langreth76}. It has been applied widely to various physical systems  \cite{Bonitz00,BS03,BF06}. A version of perturbation theory related to Schwinger-Dyson equations \cite{Landau56,LW60,Luttinger60} obeys order by order the basic conservation laws  \cite{Baym62} -- the property lost in ordinary perturbation theory. This approach allows to reproduce relatively easily the  Bogoliubov-Born-Green-Kirkwood-Yvons (BBGKY) sequence of equations \cite{Cercignani88} for correlation functions. Various applications of Keldysh technique may be found in many fields of condensed matter physics related mainly to transport phenomena including tunneling in  metal-insulator-metal junctions \cite{CCNS71} and superconductivity \cite{AG75,LO75}, see also \cite{Bonitz00,BS03,BF06,Langreth76,Danielewicz84,CSHY85,RS86,Berges04,HJ98,Rammer07}.
In high energy physics Keldysh formalism has been applied successfully to high energy scattering in QCD \cite{CGC}, to relativistic hydrodynamics \cite{hydr}, and to several other fields (see, for example, \cite{Akh} and references therein).

Wigner-Weyl calculus has been developed originally as an alternative to the operator formalism of ordinary quantum mechanics. Instead of operators in Hilbert space of functions (of coordinates or momenta) it operates with functions defined on phase space (composed of both coordinates and momenta) \cite{Wigner_1932,Groenewold_1946,Moyal_1949,Weyl}. Major motions of this formalism are Wigner distribution (to be used instead of the wave function), Weyl symbols of operators (to be used instead of the operators of the observable quantities), and Moyal product of the Weyl symbols \cite{Ali,Berezin}. Wigner-Weyl calculus has been widely applied in quantum mechanics \cite{Curtright,Zachos}. Notions of original quantum-mechanical Wigner-Weyl calculus have also been applied to solution of various problems both in condensed matter physics and in high-energy physics \cite{Cohen,Agarwal,Glauber,Husimi,Cahill,Buot}.
The Wigner distribution has been used in QCD \cite{Lorce,Eltze}. It has been considered also in quantum kinetic theory \cite{Hebenstreit,Calzetta}, and in the noncommutative field theories \cite{Bastos,Dayi}. Wigner-Weyl calculus has also been applied to the other physical problems including cosmology \cite{Habib,Chapman,Berry}.
  The notion of Wigner distribution has been used widely in the framework of Keldysh technique  \cite{Langreth76,Danielewicz84,CSHY85,RS86,Berges04,HJ98,Rammer07}.

A specific version of Wigner-Weyl formalism has been developed for quantum field theory. In this formalism the Wigner transformation of the two-point Green's function has been used, and a version of Feynmann diagram technique in terms of Wigner transformed propagators has been developed (see, for example, \cite{ZZ2019_FeynRule} and references therein). In particular, using this formalism the Hall conductivity has been represented as the topological quantity composed of Wigner-transformed Green's functions \cite{Zubkov+Wu_2019}. The similar (but simpler) constructions were used earlier to consider the intrinsic anomalous Hall effect and chiral magnetic effect \cite{Zubkov2016}.  It has been shown that the corresponding currents are proportional to  topological invariants in momentum space. This method allows to reproduce the conventional expressions  for Hall conductivity \cite{TKNN}, and to prove the absence of the equilibrium chiral magnetic effect.  In the framework of equilibrium QFT the Feynmann diagram technique that  operates with the Wigner transformed Green's functions \cite{ZZ2019_FeynRule} is useful for the investigation of non-homogeneous systems (when the two-point Green's function in momentum space depends on two momenta rather than on one).
Feynmann diagrams of this technique contain the same amount of integrations over momenta
as in the homogeneous theory (with the translational symmetry).
The price for this is the appearance of the Moyal products instead of the ordinary multiplications. However, in case of the consideration of topological quantities this is a reasonable price to be paid for the expressions that are manifestly topologically invariant. Such expressions were obtained for the QHE conductivity \cite{Zubkov+Wu_2019} and for the CSE conductivity \cite{SZ2020}. The values of these quantities are robust to smooth modifications of the systems, which allows easily to calculate their values for rather complicated systems. In particular, the QHE conductivity appears to be robust to the introduction of interactions
\cite{Zhang+Zubkov2019,Zhang_2019_JETPL}.
A version of Wigner-Weyl calculus similar to that of \cite{Zubkov2016,ZZ2019_FeynRule,Zubkov+Wu_2019} has been applied even earlier to non-equilibrium QFT i.e. to Keldysh formalism \cite{Shitade, Sugimoto,Sugimoto2006,Sugimoto2007,Sugimoto2008}. In this technique the gradient expansion has been used to calculate the fermion propagator in the way similar to that of \cite{Zubkov2016,ZZ2019_FeynRule,Zubkov+Wu_2019} (see also \cite{Polkovnikov:2009ys}). Using this technique the anomalous thermal Hall effect has been discussed as well as the QHE in certain systems. However, the consideration of Hall conductivity has been limited by the homogeneous systems. In \cite{Mokrousov,Mokrousov2} the further development of this technique was proposed that allows to deal with the non - homogeneous systems. Basing on this technique the so - called chiral Hall effect in magnets has been discussed.

In the present paper we follow the lines of research of \cite{Zubkov+Wu_2019,Zhang+Zubkov2019,Zhang_2019_JETPL,SZ2020} and \cite{Shitade,Sugimoto,Sugimoto2006,Sugimoto2007,Sugimoto2008}. We propose a method for the unification of Keldysh technique with Wigner-Weyl calculus. In order to demonstrate the power of the method we consider QHE in non-homogeneous systems, and derive rather simple expression for the conductivity through the Wigner transformed two-point Green's functions. This expression is reduced to the topological invariant of \cite{Zubkov+Wu_2019} in case of thermal equilibrium at small temperatures. It is worth mentioning that being applied to the lattice models the version of Wigner - Weyl calculus used in the present paper works for the case when inhomogeneity is negligible at the distance of the order of lattice spacing. In particular, this assumes that the values of magnetic field are much smaller than \rv{$10^4$ T}, while the wavelength of external magnetic field is much larger than $1 $ \rv{$\overcirc{\rm A}$}.

The paper is organized as follows. In Sect. \ref{SectKeldysh} we review the basis of Keldysh technique and application of Wigner-Weyl calculus to calculation of two -point Green's function in the presence of external electromagnetic field.  In Sect. \ref{SectCurrent} we present the derivation of the expression for the response of electric current to external field strength. This way we obtain the compact expression for the Hall conductivity of non-equilibrium non-homogeneous systems.     In Sect. \ref{SectEquilibrium} we demonstrate that the obtained expression for Hall conductivity is indeed reduced to the equilibrium expression of \cite{Zubkov+Wu_2019}. In Sect. \ref{SectApplication} we consider corrections to QHE conductivity of homogeneous systems from the time dependent chemical potential. General expression is derived and it is shown that this general expression gives no correction up to the first order for the system of two-dimensional Dirac fermions. In Sect. \ref{SectInvariant} we consider the case of static non-equilibrium non-homogeneous system (with interactions neglected). In Sect. \ref{SectInteraction} we discuss interaction corrections to the obtained expression. It has been shown how one loop corrections to the Hall conductivity disappear in thermal equilibrium at zero temperature. In Sect. \ref{SectConclusions} we end with conclusions.

\section{Keldysh technique for nonequilibrium systems and Wigner\,-Weyl formalism}

\label{SectKeldysh}


{We consider an inhomogeneous system interacting with external constant electric field, given by Hamiltonian $\hat\cH$. The inhomogeneity may be caused by varying external magnetic field, by elastic deformation or by other reasons. Let us consider average of an operator  $O[\psi,\bar{\psi}]$, which is a functional of field operators $\hat{\psi}, \hat{\bar{\psi}}$. Besides, we suppose that $O$ is a local quantity defined at the moment of time $t$. We have
$$
\langle O \rangle
	=  {\tr} \,\Bigl(\hat{\rho}(t_i)\, e^{- i \int_{t_i}^{t} \hat\cH dt }  O[\hat{\psi},\hat{\bar{\psi}}] e^{- i \int_{t}^{t_f} \hat\cH dt } e^{ i \int_{t_i}^{t_f} \hat\cH dt }\Bigr).
$$
Here $t_i < t < t_f$, while $\hat{\rho}(t_i)$ is density matrix at $t_i$. For an operator non-local in time we can write with help of time ordering $T$
$$
	\langle O \rangle =  {\tr} \,\Bigl(T\,\Big[\hat{\rho}(t_i)\, e^{- i \int_{t_i}^{t_f} \hat\cH dt }  O[\hat{\psi},\hat{\bar{\psi}}]\Big] e^{ i \int_{t_i}^{t_f} \hat\cH dt }\Bigr).
$$
In continuous version of Keldysh formalism an average of quantity $O$ is given by \rv{(for details see textbook \cite{Kamenev})}
$$
\langle O \rangle
	=  \int {\cal D}\bar{\psi} {\cal D} \psi\, O[\psi,\bar{\psi}]
		\exp\left\{\ii \int_C dt \int d^{\rv{D}}x\, \bar{\psi}(t,x) \hat{Q} \psi(t,x) \right\}.
$$
Here \rv{$D$ is the dimensionality of space, while}  $\psi$ and $\bar{\psi}$ are independent Grassmann variables, \rv{and by $x$ we understand here a $D$-dimensional vector.} For noninteracting condensed matter systems $\hat{Q}$ is given by $\hat{Q} = i \partial_t-\hat{H}$, where $\hat{H}$ is one-particle Hamiltonian. However, in general case of an effective description of interacting systems $\hat{Q}$ may have a more complicated form. Integration over time $t$ goes along the Keldysh contour $C$: from the initial moment of time $t_i$ to the final moment $t_f$ and back from $t_f$ to $t_i$. It is assumed that all dynamics under consideration is concentrated between these two moments of time. In particular, $O$ depends on the fields $\bar{\psi}(t,x)$ and $\psi(t,x)$ defined at $t_i < t < t_f$.
We denote the fields defined on the forward part of the contour by $\bar{\psi}_-(t,x)$ and $\psi_-(t,x)$. Those on the backward part are denoted by $\bar{\psi}_+(t,x)$ and $\psi_+(t,x)$.

While variables on the forward part of the contour are independent of those defined on the backward part, there are boundary conditions relating them to each other: $\bar{\psi}_-(t_f,x) =  \bar{\psi}_+(t_f,x)$ and $\psi_-(t_f,x)=\psi_+(t_f,x)$. At the same time integration measure ${\cal D} \bar{\psi} {\cal D} \psi$ contains integration over $\bar{\psi}_+(t_i,x)$, ${\psi}_+(t_i,x)$ and $\bar{\psi}_-(t_i,x)$, ${\psi}_-(t_i,x)$ with a weight function corresponding to a certain initial distribution given by density matrix $\hat{\rho}$. In case, when original distribution is thermal equilibrium one, we may add to the Keldysh contour after its backward piece in the complex plane of $t$ the piece that begins at $t_i$ and ends at $t_i-\ii\beta$, where $\beta = 1/T$ is inverse temperature. Fields defined along the new piece of the contour are denoted by $\bar{\psi}_\beta(t,x)$, ${\psi}_\beta(t,x)$. We supplement the integration over the fields defined on this contour by boundary conditions $\bar{\psi}_+(t_i,x) =  \bar{\psi}_\beta(t_i,x)$,  $\psi_-(t_i,x)=\psi_\beta(t_i,x)$ and $\bar{\psi}_-(t_i,x) =  \bar{\psi}_\beta(t_i-\ii\beta,x)$,  $\psi_-(t_i,x)=\psi_\beta(t_i-\ii\beta,x)$. The case of any initial density matrix $\hat{\rho}$ (not necessarily corresponding to a thermal equilibrium) may be described as follows:
\begin{eqnarray}
\langle O \rangle
&=&
	\int \frac{{\cal D}\bar{\psi}_\pm {\cal D} \psi_\pm}{{\rm Det}\, (1+\rho)} \, O[\psi_+,\bar{\psi}_+]\nonumber\\
&&
	\qquad	{\rm exp}\left\{\ii \int_{t_i}^{t_f} dt \int d^{\rv{D}}x \[\bar{\psi}_-(t,x) \hat{Q} \psi_-(t,x)-\bar{\psi}_+(t,x) \hat{Q} \psi_+(t,x)\]-\int d^{\rv{D}} x\, \bar{\psi}_-(t_i,x) {\rho} \psi_+(t_i,x)\right\} .\label{eq1}
\end{eqnarray}
Here operator $\rho$ (without hat) is related to the original density matrix of non-interacting system. Let us enumerate the eigenstates of one particle Hamiltonian $\hat{H}(t_i)$ by variable $\lambda$ (which \rv{takes} discrete values only for simplicity -- it means that we place our system \rv{in} a box with appropriate boundary conditions). Vector $|\lambda_1,\ldots, \lambda_K ; \lambda_{K+1},\ldots, \la_N\rangle$ corresponds to the occupied one-particle states $\lambda_1,\dots,\lambda_K$ and vacant one-particle states $
\lambda_{K+1},\ldots , \lambda_N$. Matrix elements of density matrix $\hat{\rho}$ between such two vectors are given by
$$
\langle \lambda_1,\ldots, \lambda_K ;\lambda_{K+1},\ldots, \lambda_N |\hat\rho| \lambda_1,\ldots, \lambda_K; \lambda_{K+1},\ldots, \lambda_N \rangle
	= \prod_{i=1...K} \frac{\langle \lambda_i |\rho|\lambda_i\rangle}{1+\langle \lambda_i |\rho|\lambda_i\rangle}
		\prod_{i=K+1,\dots,N} \frac{1}{1+\langle \lambda_i |\rho|\lambda_i\rangle}.
$$
Matrix $\rho$ is diagonal in this representation with respect to one-particle states $|\lambda\rangle$.
As a result we have, for example,
$$
{\tr} \, \hat{\rho}
	=1.
$$
\rv{Here $\frac{\langle \lambda_i |\rho|\lambda_i\rangle}{1+\langle \lambda_i |\rho|\lambda_i\rangle}$ is the probability that the one - particle state $|\lambda_i\rangle$ is occupied while  $\frac{1}{1+\langle \lambda_i |\rho|\lambda_i\rangle}$ is the probability that this state is vacant. Correspondingly, $\langle \lambda_1,\ldots \lambda_K ;\lambda_{K+1},\ldots \lambda_N |\hat\rho| \lambda_1,\ldots \lambda_K; \lambda_{K+1},\ldots \lambda_N \rangle $ is the probability that the given multi - particle system is in the state $| \lambda_1,\ldots \lambda_K; \lambda_{K+1},\ldots \lambda_N \rangle$. The sum of such probabilities over all possible multi - particle states is equal to unity as it should. }
\rv{Eq. (\ref{eq1}) is reduced to Eq. (9.14) of \cite{Kamenev} in the situation when the one - particle Hamiltonian describes a single state system. Extension to our case, when the one - particle Hamiltonian describes the system with the states numbered by quantum number $\lambda$, is obvious.}
We can also write
\begin{eqnarray}
\langle O \rangle
&=&
	\int \frac{{\cal D}\bar{\psi}_\pm {\cal D} \psi_\pm}{{\rm Det}\, (1+\rho)} \, {\rm exp}\Bigl(\ii \int_{t_i}^{t_f} dt \int d^{\rv{D}} x \[\bar{\psi}_-(t,x) \hat{Q} \psi_-(t,x)-\bar{\psi}_+(t,x) \hat{Q} \psi_+(t,x)\]\nonumber\\
&&
	\qquad
	- \int d^{\rv{D}} x \bar{\psi}_-(t_i,x) {\rho} \psi_+(t_i,x)\Bigr) O[\psi_-,\bar{\psi}_-].
\end{eqnarray}
Introducing Keldysh spinors
\be
	\Psi = \left(\begin{array}{c}\psi_-\\ \psi_+ \end{array}\right),
	\label{KelPsi}
\ee
we rewrite these expressions as follows
\begin{eqnarray}
\langle O \rangle
	&=& \frac{1}{{\rm Det}\, (1+\rho)}\int {\cal D}\bar{\Psi} {\cal D} \Psi \,
	O[\Psi,\bar\Psi]\,
	{\rm exp}\Bigl\{\ii \int_{t_i}^{t_f} dt \int d^{\rv{D}} x \bar{\Psi}(t,x) \hat{\bf Q} \Psi(t,x) \Bigr\} .
\end{eqnarray}
\rv{Again, by $x$ we understand here a $D$-dimensional vector.} Here $\bm {\hat Q}$ is in Keldysh representation
\begin{eqnarray}
	\hat{\bf Q}
	= \left(\begin{array}{cc}Q_{--} & Q_{-+}\\ Q_{+-} & Q_{++} \end{array} \right).
	\label{KelQ}
\end{eqnarray}
Naive expressions for the noninteracting particles give the following values of the components of $\hat{\bf Q}$:
\begin{eqnarray}
	Q_{++}  &=& -\hat{Q}  = -(\ii \partial_t-\hat{H}e^{+ \epsilon \partial_t}), \nonumber \\
	Q_{--}  &=&  \hat{Q}  = \ii \partial_t-\hat{H}e^{- \epsilon \partial_t}, \nonumber \\
	Q_{+-}  &=&  0, \nonumber \\
	Q_{-+}  &=& \ii \rho \delta(t-t_i)
\label{Qnaive} .
\end{eqnarray}
Here $\epsilon \to 0$ is a regularization parameter. These expressions are to be supplemented by the requirement that $\psi_-(t)$ is continued to $\psi_+(t)$ at $t=t_f$, while $\psi_+(t)$ is continued to $\psi_-(t)$ at $t=t_i$. The same refers to variables $\bar{\psi}_\pm$.
We will see below that the above naive expressions are to be corrected in order to reproduce correct expressions for the Green's  function \rv{(for details see Sect. 5.1 of \cite{Kamenev2})}. This discrepancy appears because matrix elements of operators $Q_{\alpha \beta}$ are generalized functions/distributions. Corresponding expressions for the Green's  functions are to be understood as equations for the generalized functions supplied by specific boundary conditions.

The Green's  function $\hat{\bf G}$ is defined as
\begin{eqnarray}
G_{\alpha_1 \alpha_2}(t,x|t^\prime,x^\prime)
	&=& \int \frac{{\cal D}\bar{\Psi} {\cal D} \Psi}{\ii{\rm Det}\, (1+\rho)}
	\Psi_{\alpha_1}(t,x) \bar{\Psi}_{\alpha_2}(t^\prime,x^\prime)
	\, \exp\left\{\ii \int_{t_i}^{t_f} dt \int d^{\rv{D}} x\, \bar{\Psi}(t,x) \hat{\bf Q} \Psi(t,x) \right\}.
\end{eqnarray}
Here index $\alpha$ enumerates components of $\Psi$ corresponding to the forward and backward parts of the Keldysh contour, as given by \Ref{KelPsi}. \rv{As above, by $x$ we understand here a $D$-dimensional vector.}

$\hat{\bf G}$ obeys equation
$$
\hat{\bf Q}\hat{\bf G}=1 .
$$
Notice that not all components of $\hat{\bf G}$ are independent. Namely,
\be
G_{--}+G_{++}-G_{-+}-G_{+-}=0
	\label{G-rel}
\ee
for the values of time strictly between $t_i$ and $t_f$. Correspondingly, for the values of time from the same interval we have
\be	
	Q_{--}+Q_{++}+Q_{-+}+Q_{+-}=0 .
		\label{Q-rel}
\ee
One can see, that matrix $\hat{\bf Q}$ for the non-interacting particles given by \Ref{KelQ} obeys this equation trivially. It becomes non-trivial, when we take into account interactions, which give rise to modification
$\hat{\bf Q} = \hat{\bf Q}^{(0)}-\hat{\Sigma }$, where $\hat{\Sigma}$ is the self-energy while $\hat{\bf Q}^{(0)}$ is the corresponding expression in the absence of interactions. Notice that the above requirement is broken for time equal to $t_i$ or $t_f$.

{Let us define a new set of Grassmann variables  (we keep the same notation as before, as the former will never be met again)
$$\Psi=\begin{pmatrix}\psi_1 \\ \psi_2 \end{pmatrix}$$
and
$$
	\bar{\Psi}=\begin{pmatrix}\bar{\psi_1} & \bar{\psi_2} \end{pmatrix},
$$
where $\psi_1$, $\psi_2$,$\bar{\psi}_1$, and $\bar{\psi}_2$ are related to $\psi_+$, $\psi_-$, $\bar{\psi}_+$, and $\bar{\psi}_-$  as follows:
$$
	\begin{pmatrix}\psi_1 \\ \psi_2 \end{pmatrix}=\frac{1}{\sqrt{2}}\begin{pmatrix}1 & 1 \\ 1 & -1 \end{pmatrix}\begin{pmatrix}\psi_- \\ \psi_+ \end{pmatrix}
	,\quad
	\begin{pmatrix}\bar{\psi}_1 & \bar{\psi}_2 \end{pmatrix}=\frac{1}{\sqrt{2}}\begin{pmatrix}\bar{\psi}_- & \bar{\psi}_+ \end{pmatrix}\begin{pmatrix}1 & 1 \\ -1 & 1 \end{pmatrix}.
$$
The corresponding Green's function in this representation is calculated as
\begin{eqnarray}
	\hat{\bf G}^{(K)}&=&-i\langle \begin{pmatrix}\psi_1 \\ \psi_2 \end{pmatrix}\otimes\begin{pmatrix}\bar{\psi}_1 & \bar{\psi}_2 \end{pmatrix}\rangle \\\nonumber&=&\frac{1}{2}\begin{pmatrix}1 & 1 \\ 1 & -1 \end{pmatrix}\begin{pmatrix}G^{--} & G^{-+} \\ G^{+-} & G^{++} \end{pmatrix}\begin{pmatrix}1 & 1 \\ -1 & 1 \end{pmatrix}\\\nonumber
	&=&\frac{1}{2}\begin{pmatrix}G^{--}+G^{+-}-G^{-+}-G^{++} &\quad G^{--}+G^{+-}+G^{-+}+G^{++} \\ G^{--}-G^{+-}-G^{-+}+G^{++}& \quad G^{--}-G^{+-}+G^{-+}+G^{++} \end{pmatrix}.
\label{GK}
\end{eqnarray}
Let us define Keldysh, Advanced and Retarded Green's  functions as follows:
\bes
	G^\rK  & =G^{-+}+G^{+-}=G^{--}+G^{++},\\
	G^\rA  & =G^{--}-G^{+-}=G^{-+}-G^{++},\\
	G^\rR  & =G^{--}-G^{-+}=G^{+-}-G^{++}.
\end{eqsplit}
We have
\be
	\hat{\bf G}^{(K)}=\begin{pmatrix}
	G^\rR &G^\rK \\0&G^\rA
	\end{pmatrix}.
	\label{GK}
\ee
Let us perform an additional transformation, which will lead us to the representation with the lesser component $G^<$ at the place of the Keldysh one $G^\rK $
\bes
\hat{\bf G}^{(<)}&=\begin{pmatrix}
	1&1\\0&1
	\end{pmatrix}\begin{pmatrix}
	G^\rR &G^\rK \\0&G^\rA
	\end{pmatrix}\begin{pmatrix}
	1&-1\\0&1
	\end{pmatrix}
	\\
&=	\begin{pmatrix}
	G^\rR &2G^<\\0&G^\rA
	\end{pmatrix} .
\label{G<}	
\end{eqsplit}
From this representation, an important relation follows for a product of two matrices:
\be
	(BD)^<=B^\rR D^<+B^<D^\rA .
	\label{BD<}
\ee

Now we \rv{present} the transformation rule for the inverse of the matrix Green's function $\hat{\bf Q}$ from the $(\pm)$ representation to the ``lesser" representation. From \Ref{GK} and \Ref{G<} it is evident that
\be
	\hat{\bf G}^{(<)} = U \hat{\bf G} V,
\ee
where
$$
U=\frac{1}{\sqrt{2}}\begin{pmatrix}
1&1\\0&1
\end{pmatrix}\begin{pmatrix}1 & 1 \\ 1 & -1 \end{pmatrix}=\frac{1}{\sqrt{2}}\begin{pmatrix}2 & 0\\1 &-1 \end{pmatrix}
$$
and
$$
V=\frac{1}{\sqrt{2}}\begin{pmatrix}
1&1\\-1&1
\end{pmatrix}\begin{pmatrix}1 & -1 \\ 0 & 1 \end{pmatrix}=\frac{1}{\sqrt{2}}\begin{pmatrix}1 & 0\\-1& 2 \end{pmatrix}.
$$
The corresponding inverse of $U$ and $V$ are given by
$$
U^{-1}=\frac{1}{\sqrt{2}}\begin{pmatrix}1 & 0\\1 &-2\end{pmatrix}, \qquad
	V^{-1}=\frac{1}{\sqrt{2}}\begin{pmatrix}2 & 0\\1& 1 \end{pmatrix}.
$$
$\hat{\bf G}^{(<)}$ is an upper triangle matrix in this representation, and therefore its inverse,  $\hat{\bf Q}^{(<)}$ has to be an upper triangle matrix as well. Notice, however, that this does not apply to boundary values of time $t=t_i, t_f$. We have
\begin{eqnarray}
\hat{\bf Q}^{(<)}&=&V^{-1}\hat{\bf Q} U^{-1}\\\nonumber&=&\frac{1}{2}\begin{pmatrix}2 & 0\\1& 1 \end{pmatrix}\begin{pmatrix}Q^{--} & Q^{-+} \\ Q^{+-} & Q^{++} \end{pmatrix}\begin{pmatrix}1 & 0\\1& -2 \end{pmatrix}\\\nonumber&=&
\begin{pmatrix}Q^{--}+Q^{-+} & -2Q^{-+} \\ \frac{Q^{--}+Q^{+-}+Q^{-+}+Q^{++}}{2} & -Q^{-+}-Q^{++} \end{pmatrix}\\\nonumber&=&\begin{pmatrix}Q^\rR  & 2Q^< \\ 0 & Q^\rA  \end{pmatrix},
\end{eqnarray}
where we denoted
\be
Q^\rR =Q^{--}+Q^{-+}, \qquad
	Q^\rA =-Q^{-+}-Q^{++}, \qquad
	Q^<=-Q^{-+},
\label{Qar_def}
\ee
and used \Ref{Q-rel}.

Finally, from $\hat{\bf Q}\hat{\bf G}=1$ we obtain
\begin{equation}
G^\rA  = (Q^\rA )^{-1}, \qquad G^\rR  = (Q^\rA )^{-1}, \qquad G^< =-G^\rR  Q^< G^\rA .
\label{Gar_def}
\end{equation}
Notice, that naive expressions of Eq. (\ref{Qnaive}) are to be corrected here in order to reproduce correct expressions for the Green's  functions (i.e. those obtained via discreetised approach). Inversely, the Green's functions themselves cannot be calculated inverting naive expressions of Eq. (\ref{Qnaive}). Instead the strict technique of lattice regularization is to be used or the operator formalism.

The values of $G^\rR , G^\rA , G^<$ are to be calculated using methods complement to the continuum functional integral formulation (say, lattice regularized functional integrals  \cite{Kamenev} or operator formalism). The well-known expressions for the Green's  functions of non-interacting systems that were in thermal equilibrium at $t=t_i$ are:
\bes
G^\rR
	&= (\ii \partial_t-\hat{H}e^{+ \epsilon \partial_t})^{-1}
	= (\ii \partial_t-\hat{H}+\ii \epsilon )^{-1},
		\\
G^\rA  &= ( \ii \partial_t-\hat{H}e^{- \epsilon \partial_t})^{-1}
	= ( \ii \partial_t-\hat{H}-\ii \epsilon )^{-1},
		\\
G^< &=(G^\rA -G^\rR ) \frac{\rho}{\rho+1}.
\label{Gar_expl}
\end{eqsplit}
The corresponding expressions of the elements of matrix $\hat{\bf Q}^<$ understood as the inverse to $\hat{\bf G}^<$ are:
\bes
Q^{<}&=(Q^\rA -Q^\rR )\frac{\rho}{\rho+1} = -2\ii\epsilon \frac{\rho}{\rho+1},
	\\
Q^\rR  &= \ii \partial_t-\hat{H}+\ii \epsilon ,	
	\\
Q^\rA  &=  \ii \partial_t-\hat{H}-\ii \epsilon .
\label{Qar_expl}
\end{eqsplit}
Being understood formally, expression for $Q^<$ tends to zero at $\epsilon\to 0$. However, the small but nonzero value of $\epsilon$ is to be kept in order to calculate $\hat{\bf G}^<$ which is a generalized rather than an ordinary function.

Notice, that  in general case functions $G^\rR ,G^\rA ,G^<$ obey important constraints
\be
[G^\rR ]^+ = G^\rA , \quad [G^<]^+ =-G^<.
\label{G-const}
\ee
Together with \Ref{Gar_def} it means that $\ii Q^<$ remains Hermitian both for the case of non-interacting equilibrium systems, and for the more complicated ones.

\rv{We would like to stress that up to now we presented the pedagogical introduction to Keldysh formalism based on the textbook materials given in \cite{Kamenev,Kamenev2}. Below we remind the readers the basic notions of Wigner - Weyl calculus (see, for example, \cite{Zubkov+Wu_2019} or \cite{Sugimoto}). } 
 \rv{By large latine letters we denote the $D+1$ dimensional vectors containing both space and time components. For matrix elements  $A(X_1,X_2) = \langle X_1 | \hat{A} | X_2 \rangle $ } of operator $\hat{A}$  \rv{the Wigner transformation} is defined as 
\rv{\be
A_W(X|P)=\int d^{D+1} Y\, e^{\ii Y^\mu P_\mu }A(X+Y/2,X-Y/2),\quad \mu =0,1,...,D-1
	\label{WignerTr}
\ee
As usual $P^\mu=(P^0,p)$ while $P_\mu = (P^0,-p)$, where $p$ is the $D$ - dimensional spatial momentum.}
Then the inverse transformation gives
\rv{$$
A(X+Y/2,X-Y/2)|_{Y\rightarrow +0} =\int\frac{d^{D+1}P}{(2\pi)^3}A_W(X|P).
$$}
The Wigner symbol of $\hat{\bf G}$ will be denoted by $\hat{G}$ while Wigner transformation of $\hat{\bf Q}$  -- by $\hat{Q}$.  The components of the matrices will not be supplied with the subscript $W$ indicating whether it is Wigner-transformed quantity or not. It will always be clear from the context if we mean Wigner-transformed $Q$ and $G$, or their original operator form. The Wigner transformed functions obey Groenewold equation
\begin{equation}
    \hat{Q} * \hat{G} = 1.
\end{equation}
The star (Moyal) product $*$ is defined as
\begin{equation}
  \left(A* B\right)(X|P) = A(X|P)\,e^{\rv{-}\ii(\overleftarrow{\partial}_{X^{\mu}}\overrightarrow{\partial}_{P_{\mu}}-\overleftarrow{\partial}_{P_{\mu}}\overrightarrow{\partial}_{X^{\mu}})/2}B(X|P).
\end{equation}
In the presence of interactions $\hat{Q} = \hat{Q}^{(0)}(X|\pi)-\hat{\Sigma }(X|\pi)$, where $\hat{\Sigma}$ is the Wigner transformed self-energy while $\hat{Q}^{(0)}$ is the corresponding expression in the absence of interactions.\textbf{}

Further we will consider the case, when external electromagnetic field $A$ with constant field strength ${\cal F}^{\mu\nu}$ is added. At the same time there may be other gauge fields that are allowed to vary both in space and time. External field strength ${\cal F}^{\mu \nu}$ is assumed to be small. To leading order in the field strength, the introduction of such a field results in the substitution $P \to \pi = P- A$.  \rv{(In our notations $\pi^\mu$ is $D+1$ - dimensional vector like $P^\mu$, and its spatial components change sign when the index is lowered.)} Without loss of generality we can decompose the $*$-product as
\begin{equation}
    * =  \star~ e^{\rv{-}\ii  \mathcal{F}^{\mu\nu}\overleftarrow{\partial}_{\pi^{\mu}}\overrightarrow{\partial}_{\pi^{\nu}}/2}.
\end{equation}
Here
\begin{equation}
  \left(A\star B\right)(X|\pi) = A(X|\pi)\,e^{\rv{-\ii(\overleftarrow{\partial}_{X^{\mu}}\overrightarrow{\partial}_{\pi_{\mu}}-\overleftarrow{\partial}_{\pi_{\mu}}\overrightarrow{\partial}_{X^{\mu}})/2}}B(X|\pi).
\end{equation}
Let us consider the perturbative expansion of $\hat{Q}$ and $\hat{G}$ up to the first order in $\mathcal{F}^{\mu\nu}$
\begin{equation}
    \hat{Q} = \hat{Q}  +\frac{1}{2}\mathcal{F}^{\mu\nu}\hat{Q}_{\mu\nu}^{(1)},\quad
    \hat{G} = \hat{G}  +\frac{1}{2}\mathcal{F}^{\mu\nu}\hat{G}_{\mu\nu}^{(1)}.\label{QGK}
\end{equation}
The Groenewold equation reads
\begin{equation}
  \left(\hat{Q}  +\frac{1}{2}\mathcal{F}^{\mu\nu}\hat{Q}_{\mu\nu}^{(1)}\right)\star~ e^{\rv{-}i  \mathcal{F}^{\mu\nu}\overleftarrow{\partial}_{\pi^{\mu}}\overrightarrow{\partial}_{\pi^{\nu}}/2}\left(\hat{G}  +\frac{1}{2}\mathcal{F}^{\mu\nu}\hat{G}_{\mu\nu}^{(1)}\right) = 1.
  \label{Groe-F}
\end{equation}
In the zeroth order in $\cal F$ we have $\hat{Q} \star \hat{G}  = 1$, and
$\hat{Q} \star \hat{G}^{(1)}+\hat{Q}^{(1)}\star\hat{G} \rv{-} \ii \hat{Q} \star \overleftarrow{\partial}_{\pi^{\mu}}\overrightarrow{\partial}_{\pi^{\nu}} \hat{G}  = 0$ in the first one. Then we can write
\begin{equation}
     \hat{G}_{\mu\nu}^{(1)} =-\hat{G} \star  \hat{Q}_{\mu\nu}^{(1)}\star \hat{G}   \rv{-}   \ii\left(\hat{G} \star \partial_{\pi^{\mu}}\hat{Q}  \star\hat{G} \star \partial_{\pi^{\nu}}\hat{Q} \star \hat{G} -(\mu\leftrightarrow \nu)\right)/{2}
     .
     \label{QGK1}
\end{equation}

\section{Derivation of current density }

\label{SectCurrent}

\rv{Our derivation in this section closely follow those of \cite{Sugimoto}.} In the non-interacting theory operator of electric current density is given by
$$
	\hat{j^i}=-\hat{\bar{\psi}} \frac{\partial \hat{Q}}{\partial p_i} \hat{\psi},\quad
		i=1,2,\ldots D.
$$
\rv{Notice that here $p_i = P^i = - P_i$ are the spatial components of momentum. It is a vector in Euclidean $D$ - dimensional space, and the position (upper or lower) of index $i$ is irrelevant: $p_i = p^i$.} Therefore, in nonequilibrium non-interacting models the average current density depending on time may be calculated as follows:
\bes
\langle j^i(t,x) \rangle
	&\equiv  \int \frac{{\cal D}\bar{\psi}_\pm {\cal D} \psi_\pm}{{\rm Det}\, (1+\rho)} \,
	\(- \bar{\psi}_-(t,x) \partial_{p_i} \hat{Q} \psi_-(t,x) \)
	\, \exp\left\{\ii \int_{t_i}^{t_f} dt \int d^{\rv{D}} x\, \bar{\Psi}(t,x) \hat{\bf Q} \Psi(t,x) \right\}
	\\
&= \int \frac{{\cal D}\bar{\psi}_\pm {\cal D} \psi_\pm}{{\rm Det}\, (1+\rho)} \,
	\(- \bar{\psi}_+(t,x) \partial_{p_i} \hat{Q} \psi_+(t,x) \)
	\, \exp\left\{\ii \int_{t_i}^{t_f} dt \int d^{\rv{D}}x\, \bar{\Psi}(t,x) \hat{\bf Q} \Psi(t,x) \right\}
	\\
&=
\frac{1}{2}\int \frac{{\cal D}\bar{\psi}_\pm {\cal D} \psi_\pm}{{\rm Det}\, (1+\rho)} \,
	\(- \bar{\psi}_+(t,x) \partial_{p_i} \hat{Q} \psi_+(t,x) )- \bar{\psi}_-(t,x) \partial_{p_i} \hat{Q} \psi_-(t,x) \)
	\, \exp\left\{\ii \int_{t_i}^{t_f}\!\! dt \int d^{\rv{D}} x\, \bar{\Psi}(t,x) \hat{\bf Q} \Psi(t,x) \right\} .
\end{eqsplit}
We rewrite this expression as follows
\begin{eqnarray}
\langle j^i(t,x) \rangle
	&=&-\frac{\ii}{2}{\tr} \[\hat{\bf G} \hat{{\bf v}}^i\].
\end{eqnarray}
Here the velocity operator is defined as
$$
\hat{\bf v}^i = \partial_{p_i}\begin{pmatrix}-Q^{--} & 0 \\ 0 & Q^{++} \end{pmatrix}.
$$
In the ``lesser" representation it has the following form:
\begin{eqnarray}
\hat{\bf v}_i^{(<)}&=&\partial_{p_i}\frac{1}{2}\begin{pmatrix}2 & 0\\1& 1 \end{pmatrix}\begin{pmatrix}-Q^{--} & 0 \\ 0 & Q^{++} \end{pmatrix}\begin{pmatrix}1 & 0\\1& -2 \end{pmatrix}\\\nonumber
&=&\partial_{p_i}\begin{pmatrix} -Q^{--} & 0 \\ \frac{-Q^{--}+Q^{++}}{2}  & -Q^{++} \end{pmatrix}\\\nonumber
&=&\partial_{p_i}\begin{pmatrix} -Q^\rR -Q^< & 0 \\ -\frac{Q^\rR +Q^\rA }{2}  & -Q^<+Q^\rA  \end{pmatrix},
\end{eqnarray}
where we have used that according to \Ref{Qar_def} $Q^{--}=Q^\rR +Q^<$, $Q^{-+}=-Q^<$, $Q^{+-}=-Q^\rR +Q^\rA -Q^<$, and $Q^{++}=Q^<-Q^\rA $.
For the current density we have
\begin{eqnarray}
\langle j^i \rangle
&=&- \frac{\ii}{2}\tr\left[\hat{\bf G}\hat{\bf v}^i\right]
	=-\frac{\ii}{2}\tr\left[\begin{pmatrix}
	G^\rR &2G^<\\0&G^\rA
	\end{pmatrix}\partial_{p_i}\begin{pmatrix}- Q^\rR -Q^< & 0 \\ -\frac{Q^\rR +Q^\rA }{2}  & -Q^<+Q^\rA  \end{pmatrix}\right]\\\nonumber
&=&\frac{\ii}{2}\Tr \left(G^\rR \partial_{p_i} Q^\rR -G^\rA \partial_{p_i}Q^\rA \right)
	+\frac{\ii}{2}\Tr \left(G^\rR \partial_{p_i} Q^<+G^<\partial_{p_i} Q^\rA \right)
		+\frac{\ii}{2}\Tr \left(G^\rA \partial_{p_i}Q^< +G^<\partial_{p_i} Q^\rR \right)
		\\ \nonumber
&=& \rm I+\rm II+\rm III.
\end{eqnarray}
Here we omit the trace over the internal indices for simplicity of notations.
The second term may be represented as $\rm II =\frac{\ii}{2}\Tr\left({G}\partial_{p_i} {Q}\right)^<$ by using \Ref{BD<}.

Basing on \Ref{G-const} we come to the conclusion that the third term $\rm III$ is complex conjugate of $\rm II$. Therefore, we  can write
\be
\langle j^i \rangle
	=\frac{\ii}{2}\Tr\left(\hat{\bf G}\partial_{p_i}\hat {\bf Q}\right)^\rR
	+ \frac{\ii}{2}\Tr\left(\hat{\bf G}\partial_{p_i}\hat {\bf Q}\right)^<+{\rm c.c.}
\label{<j>}
\ee
It is easy to express the current density in terms of the Wigner transformations of $\hat{\bf G}$ and $\hat{\bf Q}$, see \Ref{WignerTr}. For the sufficiently smooth external fields we have
\bes
J^i(X) \equiv \langle j^i(t,x) \rangle
& 	= \rv{-} \frac{\ii }{2}\int \frac{d^{\rv{D+1}}\pi}{(2\pi)^{\rv{D+1}}}
		\tr\left(\hat{G} (\partial_{\pi_{i}}\hat{Q})\right)^{\rR}
	\rv{-} \frac{\ii }{2}\int \frac{d^{\rv{D+1}}\pi}{(2\pi)^{\rv{D+1}}}
			\tr\left(\hat{G} (\partial_{\pi_{i}}\hat{Q})\right)^{\rA}	
		\\
&\qquad		
	\rv{-}\frac{\ii }{2}\int \frac{d^{\rv{D+1}}\pi}{(2\pi)^{\rv{D+1}}}
		\tr\left(\hat{G} (\partial_{\pi_{i}}\hat{Q})\right)^{<}
			\rv{-}\frac{\ii }{2}\int \frac{d^{\rv{D+1}}\pi}{(2\pi)^{\rv{D+1}}}
				\tr\left((\partial_{\pi_{i}}\hat{Q}) \hat{G}\right)^{<} .
\end{eqsplit}
Here \rv{$D=2$} for the $2+1$ dimensional systems. \rv{Notice, that $\pi_i$ is $D+1$ dimensional vector, and $\pi_i = - \pi^i = - p^i = - p_i$ for $i = 1,..., D$.}
The vanishing imaginary part $\pm \ii \epsilon$ in \Ref{Gar_expl} indicates that all possible poles of $G^{\rR}$  ($G^{\rA}$) lie in the lower/upper complex half-plane in $\omega\equiv \pi_0$. Then, the contour of integration over $\pi_0$ in the first term of the first line of the above can be closed in the upper half plane, where the integrand does not have poles, giving exactly zero. Integral in the second term is similarly equal to zero because now the contour can be closed in the lower half plane.  We are left with
\be
J^i(X) = \rv{-}
	\frac{\ii }{2}\int \frac{d^{\rv{D+1}}\pi}{(2\pi)^{\rv{D+1}}} \tr\left((\partial_{\pi_{i}}\hat{Q})\hat{G}\right)^{<}
	\rv{-}\frac{\ii }{2}\int \frac{d^{\rv{D+1}}\pi}{(2\pi)^{\rv{D+1}}}
		\tr\left(\hat{G} (\partial_{\pi_{i}}\hat{Q})\right)^{<}.
\label{J Wigner}
\ee
Using expansion of Eqs. (\ref{QGK})-(\ref{QGK1}) we obtain the following expression for the linear response of the electric current to constant external field strength $\mathcal{F}^{\mu\nu}$:
\begin{eqnarray}
{J}^i
	&=&    -\frac{1}{4}\int \frac{d^{\rv{D+1}}\pi}{(2\pi)^{\rv{D+1}} }  \tr\Bigl(\hat{G} \star \partial_{\pi^{\mu}}\hat{Q}  \star\hat{G} \star \partial_{\pi^{\nu}}\hat{Q} \star \hat{G}  \partial_{\pi_{i}}\hat{Q} \Bigr)^{<}\mathcal{F}^{\mu\nu}\nonumber\\
&&
	-\frac{1}{4}\int \frac{d^{\rv{D+1}}\pi}{(2\pi)^{\rv{D+1}}}  \tr\Bigl(\partial_{\pi_{i}}\hat{Q}  \hat{G} \star \partial_{\pi^{\mu}}\hat{Q}  \star\hat{G} \star \partial_{\pi^{\nu}}\hat{Q} \star \hat{G}  \Bigr)^{<}\mathcal{F}^{\mu\nu}.
\end{eqnarray}

Assuming that $\cal F$ includes only electric field, we represent this expression in two-dimensional systems as:
$$
	{J}^i = \sigma^{ij}  \mathcal{F}_{0j} ,
$$
where the conductivity tensor $\sigma^{ij}$ may be given as follows:
\begin{equation}
    \sigma^{ij} =  \frac{1}{4} \int \frac{d^{\rv{D+1}}\pi}{(2\pi)^{\rv{D+1}} } \tr\left(\partial_{\pi_{i}}\hat{Q}  \left[\hat{G} \star \partial_{\rv{\pi_{[0}}}\hat{Q}  \star \partial_{\rv{\pi_{j]}}}\hat{G}  \right]\right)^< +{\rm c.c.}\label{MAIN}
\end{equation}
Here $(...)_{[0} (...)_{ j]} =(...)_{0} (...)_{ j} -(...)_{j} (...)_{ 0}  $ means anti-symmetrization.
In the most general case, conductivity may be expressed as a sum of symmetric and anti-symmetric parts
$\sigma^{ij} = \sigma^{ij}_{H} + \sigma^{ij}_S$. Here $\sigma^{ij}_H = (\sigma^{ij}-\sigma^{ji})/2$ is the Hall part of conductivity while $\sigma^{ij}_S = (\sigma^{ij}+\sigma^{ji})/2$ is the conventional symmetric conductivity.

\section{Equilibrium limit of Hall conductivity}
\label{SectEquilibrium}
In order to compare the results obtained above with those of equilibrium theory here we derive the equilibrium limit of the Hall conductivity for the system in consideration. Let us start by recalling the main result for the conductivity out of equilibrium \Ref{MAIN}. Upon averaging the conductivity over the whole volume $\cal V$ of the system and over the overall time of the process, we can restore the missing $\star$ product (see \cite{FZ2019_2} for details of the Wigner-Weyl calculus in general)
\be
\bar \sigma^{ij}
	=  \rv{-}\frac{1}{4} \int \frac{d^{\rv{D+1}}\pi d^{\rv{D+1}} X}{(2\pi)^{\rv{D+1}}  \, \beta\,{\cal V} } 	
		\tr\left(\partial_{\pi_{i}}\hat{Q} \star \hat{G} \star \partial_{\pi_{[0}}\hat{Q}  \star \hat{G}\star \partial_{\pi_{j]}}\hat{Q}\star\hat{G}\right)^< +{\rm c.c.}
\ee
and the \rv{Wick rotation is possible (see  Appendix \ref{App 1a})}:
\be
\bar \sigma^{ij}
	= \rv{-} \frac{1}{4} \int \frac{d^{\rv{D+1}}\Pi d^{\rv{D+1}} X}{(2\pi)^{\rv{D+1}}  \, \beta\,{\cal V}} 	
		\tr\left(\partial_{\rv{\Pi_{i}}}\hat{Q}^\rM  \star \hat{G}^\rM  \star \partial_{\rv{\Pi_{[D+1}}}\hat{Q}^\rM   \star \hat{G}^\rM \star \partial_{\rv{\Pi_{j]}}}\hat{Q}^\rM \star\hat{G}^\rM \right).
\ee
\rv{Here $\Pi$ is Euclidean $D+1$ - momentum, i.e. $\Pi^{D+1} = \omega$ is continuous Matsubara frequency, while $\Pi^i = \pi^i$ for $i=1,...,D$. Notice that $\partial_{\pi^0} = -\ii \partial_{\Pi^{D+1}}$, the superscript M indicates that quantities are taken at Matsubara frequency.}
Thus we obtain for the conductivity of $2+1$ dimensional system averaged over the whole of its volume and over the overall time of the process:
$$
 \bar{\sigma}^{ij}  = \frac{{\cal N}}{2\pi}\epsilon^{ij},
$$
where
\begin{eqnarray}
   {\cal N} &=&\rv{+}2\pi  \, T \, \frac{1}{24\pi^2 \, \beta\,{\cal V}}\epsilon^{\mu\nu\rho} \int d^3 X \sum_{\omega_n = 2\pi T(n+1/2)}\int d^2\Pi
   \tr\left(\partial_{\rv{\Pi^{\mu}}}\hat{Q} ^\rM  \star\hat{G} ^\rM \star \partial_{\rv{\Pi^{\nu}}}\hat{Q} ^\rM  \star\hat{G} ^\rM \star \partial_{\rv{\Pi^{\rho}}}\hat{Q} ^\rM \star \hat{G} ^\rM \right).
\end{eqnarray}
Here the sum is over Matsubara frequencies, and we substitute $\pi_0$ in the argument of $Q$ and $G$  by $\ii\omega_n$, $\epsilon^{ij}$ and $\epsilon^{\mu\nu\rho}$  are totally antisymmetric tensors.

In the required limit of small temperatures the sum over Matsubara frequencies is reduced to an integral and we arrive at the needed expression:
  \begin{eqnarray}
   {\cal N} &=&\rv{+} \frac{1}{24\pi^2 \, \beta\,{\cal V}}\epsilon^{\mu\nu\rho} \int d^3 X \int d^3\Pi
   \tr\left(\partial_{\rv{\Pi^{\mu}}}\hat{Q} ^\rM  \star\hat{G} ^\rM \star \partial_{\rv{\Pi^{\nu}}}\hat{Q} ^\rM  \star\hat{G} ^\rM \star \partial_{\rv{\Pi^{\rho}}}\hat{Q} ^\rM \star \hat{G} ^\rM \right).
   \label{NEQ}
\end{eqnarray}
Here ${\cal V}$ is the system volume.
One can check that this expression is a topological invariant. For that we need that the system was in thermal equilibrium originally, and that the thermal equilibrium corresponds to zero temperature. Moreover, we need that the Hamiltonian does not depend on time, which means that the systems remains in thermal equilibrium during the whole process. The value of the average conductivity $\bar{\sigma}^{ij}$ is then robust to smooth variations of the system. (This does not refer, however, to local values of conductivity that may depend on space coordinates.) Notice, that the integration over $x$ is important for the topological invariance of this quantity. Below we will see, that it is equal to the topological invariant obtained using the conventional equilibrium technique.

Finally, a few words on the relation between the Wigner transformation of the Matsubara Green's function ($G^\rM $) and the retarded (advanced) Green's function $G^\rR  (G^\rA )$ from above are in place. It may be seen through the way the poles in the complex plane of $\omega$ are surrounded by the deformed integration contour. 

\rv{Recall that the equilibrium theory appears in the description of the system with the Hamiltonian $\hat{H}$ that does not depend on time. Let us denote by $G$ the Green's  function, in which the poles in $\omega$ give true singularities, i.e. 
$$
G(x_1,x_2,\omega) \equiv \langle x_1|(\omega - \hat{H})^{-1}|x_2\rangle
$$
 Then the time ordered Green's  function (Feynmann propagator) is defined as}
 \begin{equation}
 G^\rT ( x, x^{\prime}, \omega)
 	= \lim\limits_{\eta\rightarrow 0} G(x, x^{\prime}, \omega+\ii\eta \, {\rm sign} \,\omega).
 \end{equation}
The retarded Green's  function is defined as
\begin{equation}
 G^\rR ( x, x^{\prime}, \omega) = \lim\limits_{\eta\rightarrow 0} G(x, x^{\prime}, \omega+\ii\eta),
\end{equation}
and similarly for the advanced Green's function we have
\begin{equation}
  G^\rA ( x, x^{\prime}, \omega) = \lim\limits_{\eta\rightarrow 0} G(x, x^{\prime}, \omega-\ii\eta).
\end{equation}
The Matsubara Green's function $G^\rM $ is defined as
\be
	G^\rM ( x, x^{\prime},\omega_n) = G( x, x^{\prime},\ii\omega_n),
	\label{Mats}
\ee
or in terms of imaginary time $\tau$:
\begin{equation}
G^\rM ( x, x^{\prime},\tau)
  = \frac{1}{\beta}\sum\limits_{n=-\infty}^{\infty}e^{-\ii\omega_n \tau}G( x, x^{\prime},\ii\omega_n).
\end{equation}
Here $\omega_n = (2n+1)\pi/{\beta}$ is the Matsubara frequency while $\beta$ is the inverse of temperature.

The above relations between the retarded (advanced)  and Matsubara Green's functions  are valid in the Fourier transformed domain ($\omega$ space). \rv{They may be extended to the Wigner - transformed Green functions as follows. For the retarded Green function we have 
\begin{equation}\label{Retarded Green's  Wigner}
  G^\rR _W({x},{p},T,\omega) = {\rm lim}_{\eta\to +0}\int d^{D} y\, e^{-\ii y p}G(x+y/2,x-y/2,\omega + i \eta),
\end{equation}
The similar expressions are valid for advanced and Matsubara Green functions: 
\begin{equation}\label{Retarded Green's  Wigner}
	G^\rA _W({x},{p},T,\omega) = {\rm lim}_{\eta\to +0}\int d^{D} y\, e^{-\ii y p}G(x+y/2,x-y/2,\omega - i \eta),
\end{equation}
and 
\begin{equation}\label{Retarded Green's  Wigner}
	G^\rM _W({x},{p},T,\omega) = {\rm lim}_{\eta\to +0}\int d^{D} y\, e^{-\ii y p}G(x+y/2,x-y/2,i\omega )
\end{equation} 
One can see that those Wigner - transformed Green functions do not depend on time $T$. Such a dependence is back for the case of the non - equilibrium systems. In the latter case, however, the Matsubara Green function is not defined.}

\section{Corrections to Hall conductivity due to the time dependent chemical potential }

\label{SectApplication}

\subsection{Time dependent perturbation of one-particle Hamiltonian}
In this section we consider evolution of $2+1 $ dimensional  equilibrium homogeneous system of non-interacting electrons subject to a time dependent perturbation.  Hall conductivity of non-perturbed system is given by \cite{Volovik2003}
$$
\sigma_H = \rv{-}\frac{{\cal N}}{2\pi},
$$
where $\cal N$ is the topological quantity of Eq. (\ref{NEQ}), applied in this case
\be
{\cal N}
   =\rv{+} \frac{1}{24\pi^2 \, }\epsilon^{\mu\nu\rho} \int d^3\rv{\Pi}
   	\tr\left(
   		\partial_{\rv{\Pi^{\mu}}}\hat{Q}_0 ^\rM  \hat{G}_0^\rM  \partial_{\Pi^{\nu}}\hat{Q}_0^\rM  \hat{G}_0^\rM  \partial_{\Pi^{\rho}}\hat{Q}_0^\rM  \hat{G}_0^\rM
   	\right).
\label{NEQ0}
\ee
We introduced here a subscript $0$ to indicate non-perturbed system. The star is omitted here because for homogeneous time-independent systems the Wigner transformation is equal to the Fourier one, and thus Matsubara Green's function does not contain dependence on coordinates, nor on time.

We will be interested in consideration of the time dependent perturbation of the following form:
\begin{equation}
\hat{Q} =\hat{Q}_0 + \mu_0 \cos \omega_0 t,
\qquad \hat{Q}_0\equiv \omega-\hat{H} .
\end{equation}
Here $\hat{H}$ is the original one-particle Hamiltonian of the unperturbed system. It is supposed that the system is homogeneous, i.e. in the present section we are speaking of the intrinsic anomalous quantum Hall effect. Therefore, in the absence of external electric field $\hat{H} = H(\hat{p})$, where $\hat{p}$ is operator of momentum. For concreteness we may take the Haldane model or any other model with intrinsic AQHE.

Using perturbation theory in powers of $\mu_0$ we are going to calculate a correction to equilibrium topological conductivity proportional to $\mu_0$. It is supposed, that initially the system was in the thermal equilibrium at zero temperature.

In the general case the conductivity tensor $\sigma^{ij}$ is given by \Ref{MAIN}
\be
\sigma^{ij}
=
\rv{-}	\frac{1}{4} \int \frac{d^{D+1}\pi}{(2\pi)^{D+1} } \tr\left(\partial_{\pi^{i}}\hat{Q}
		\left[
			\hat{G} \star \partial_{\pi^{[0}}\hat{Q}  \star \hat{G}\star \partial_{\pi^{j]}}\hat{Q} \star  \hat{G}
		\right]\right)^<
\rv{-}
	\frac{1}{4} \int \frac{d^{D+1}\pi}{(2\pi)^{D+1} } \tr\left(
		\left[
			\hat{G} \star \partial_{\pi^{[0}}\hat{Q}  \star \hat{G}\star \partial_{\pi^{j]}}\hat{Q} \star  \hat{G}
		\right]
		\partial_{\pi^{i}}\hat{Q} \right)^< .
\ee
This sum is real for any $\hat Q$, and thus we can trivially rewrite it as follows
\be
\sigma^{ij}
=
	\rv{-}\underbrace{\frac{1}{8} \int \frac{d^{D+1}\pi}{(2\pi)^{D+1} } \tr\left(\partial_{\pi^{i}}\hat{Q}
		\left[
			\hat{G} \star \partial_{\pi^{[0}}\hat{Q}  \star \hat{G}\star \partial_{\pi^{j]}}\hat{Q} \star  \hat{G}
		\right]\right)^< }_{\bf I}
\rv{-}
	\underbrace{\frac{1}{8} \int \frac{d^{D+1}\pi}{(2\pi)^{D+1} } \tr\left(
		\left[
			\hat{G} \star \partial_{\pi^{[0}}\hat{Q}  \star \hat{G}\star \partial_{\pi^{j]}}\hat{Q} \star  \hat{G}
		\right]
		\partial_{\pi^{i}}\hat{Q} \right)^<}_{\bf II}
+ {\rm c.c.}
\label{si_real}
\ee
Now, up to the terms linear in $\mu_0$ we can formally substitute in the above
\begin{equation}
Q^{R} \to \tilde Q^{R} = \omega- H(\pi) + \mu_0 {\rm exp} (\ii\omega_0 t)+\ii\epsilon,
	\qquad
	Q^{A} \to \tilde  Q^{A} = \omega-H(\pi) + \mu_0 {\rm exp} (\ii\omega_0 t)-\ii\epsilon
\end{equation}
while the ``lesser" component is the same
\begin{equation}
	Q^< =- 2 \ii \epsilon f(\hat\pi^{{0}}),
\end{equation}
and $Q\star G = 1$. Note that we use this substitution only formally, as in such notation $[\tilde Q^\rA ]^+\ne \tilde Q^\rR $.

For the variation of the conductivity, $\Delta \sigma = \sigma(\mu_0)-\sigma(0)$,  we will have the following contribution corresponding to $\bf I$ of \Ref{si_real}:
\bes
\Delta^{\bf I}\sigma^{ij}
&=
\rv{-}	\frac{1}{8} \int \frac{d^{D+1}\pi}{(2\pi)^{D+1} } \tr
		\left(\partial_{\pi^{i}}\hat{Q}_0  \left[\Delta\hat{G} \star \partial_{\pi^{[0}}\hat{Q}_0  \star \hat{G}_0\star \partial_{\pi^{j]}}\hat{Q}_0\star\hat{G}_0  \right]
		\right)^< \\
& \qquad		
	\rv{-} 	\frac{1}{8} \int \frac{d^{D+1}\pi}{(2\pi)^{D+1} } \tr
			\left(\partial_{\pi^{i}}\hat{Q}_0  \left[\hat{G}_0 \star \partial_{\pi^{[0}}\hat{Q}_0  \star \Delta\hat{G}\star \partial_{\pi^{j]}}\hat{Q}_0\star\hat{G}_0  \right]
			\right)^<\\
& \qquad		
	\rv{-} 	\frac{1}{8} \int \frac{d^{D+1}\pi}{(2\pi)^{D+1} } \tr
			\left(\partial_{\pi^{i}}\hat{Q}_0  \left[\hat{G}_0 \star \partial_{\pi^{[0}}\hat{Q}_0 \star \hat{G}_0\star \partial_{\pi^{j]}}\hat{Q}_0 \star\Delta\hat{G}  \right]
			\right)^<,
	\label{delta si}
\end{eqsplit}
The contribution $\Delta^{\bf II}\sigma^{ij}$ coming from $\bf II$ of \Ref{si_real}, is completely similar, but with $\partial_{\pi^{i}}\hat{Q}_0$ interchanged position with the square bracket in the traces. Here we denoted
\be
\Delta\hat G =-\hat G_0\star \Delta \hat Q\star \hat G_0,
	\qquad
	\Delta \hat Q = \mu_0\, e^{\ii \omega_0 t}
		\begin{pmatrix}
		1 &0\\
		0&1
		\end{pmatrix}
		.
\ee
Note that in \Ref{delta si} we do not need to consider $\Delta \hat Q$-s directly as they always are subject to differentiation in $\pi$ and thus do not contribute.

To be able to calculate \Ref{delta si} explicitly, let us derive a useful identity for the star product containing $e^{\ii\omega_0 t}$
\begin{eqnarray}
{\rm exp} (\ii\omega_0 t) \star h(\omega)
	&=&
		{\rm exp} (\ii\omega_0 t) e^{\ii\overleftarrow{\partial_t} \partial_\omega/2} h(\omega) =  {\rm exp} (\ii\omega_0 t) e^{-\omega_0 \partial_\omega/2} h(\omega) =  {\rm exp} (\ii\omega_0 t) h(\omega-\omega_0/2).
\label{[+/-]}
\end{eqnarray}
Then,
\be
	\Delta\hat G =-\mu_0\, e^{\ii \omega_0 t} \hat G_0(\omega+\omega_0/2) \hat G_0(\omega-\omega_0/2).
\ee
Recall, that $\hat G_0$ does not depend neither on time, nor coordinate. Therefore, using again  \Ref{[+/-]} (and its complex conjugate for terms of the type $ h(\omega) \star e^{\ii\omega_0 t}$) we obtain
\bes
\Delta^{\bf I}\sigma^{ij}
&=
	\frac{\mu_0\, e^{\ii \omega_0 t} }{8} \int \frac{d^{D+1}\pi}{(2\pi)^{D+1} } \tr
		\left(\partial_{\pi^{i}}\hat{Q}_0^{[0]}  \left[\hat{G}^{[+]}   \hat{G}^{[-]}   \partial_{\pi^{[0}}\hat{Q}_0^{[-]}    \hat{G}_0^{[-]}  \partial_{\pi^{j]}}\hat{Q}_0^{[-]} \hat{G}_0^{[-]}  \right]
		\right)^< \\
& \qquad		
	\rv{+} 	\frac{\mu_0\, e^{\ii \omega_0 t} }{8} \int \frac{d^{D+1}\pi}{(2\pi)^{D+1} } \tr
			\left(\partial_{\pi^{i}}\hat{Q}_0^{[0]}  \left[\hat{G}_0^{[+]}   \partial_{\pi^{[0}}\hat{Q}_0^{[+]}    \hat{G}^{[+]}  \hat{G}^{[-]}  \partial_{\pi^{j]}}\hat{Q}_0^{[-]} \hat{G}_0^{[-]}  \right]
			\right)^<\\
& \qquad		
	\rv{+} 	\frac{\mu_0\, e^{\ii \omega_0 t} }{8} \int \frac{d^{D+1}\pi}{(2\pi)^{D+1} } \tr
			\left(\partial_{\pi^{i}}\hat{Q}_0^{[0]}  \left[\hat{G}_0^{[+]}   \partial_{\pi^{[0}}\hat{Q}_0^{[+]}   \hat{G}_0^{[+]}  \partial_{\pi^{j]}}\hat{Q}_0^{[+]}  \hat{G}^{[+]} \hat{G}^{[-]}  \right]
			\right)^<,
	\label{delta si+}
\end{eqsplit}
where we introduced superscript to denote the shifting of the frequency
\be
	K^{[\pm]} \equiv K(\om\pm\om_0/2),\qquad
		K^{[0]} \equiv K(\om).
\label{[pm]}
\ee

Now we are in a position to apply the results of Appendix \ref{App 1a}, and state that in the limit of zero initial temperature
\bes
\Delta^{\bf I}\sigma^{ij}
&=
	\frac{\mu_0\, e^{\ii \omega_0 t} }{8} \int \frac{d^{D+1}{\rv{\Pi}}}{(2{\rv{\pi}})^{D+1} } \tr
		\left(\partial_{{\rv{\Pi}}^{i}}\hat{Q}_0^{[0]}  \left[\hat{G}^{[+]}   \hat{G}^{[-]}   \partial_{{\rv{\Pi}}^{[0}}\hat{Q}_0^{[-]}    \hat{G}_0^{[-]}  \partial_{{\rv{\Pi}}^{j]}}\hat{Q}_0^{[-]} \hat{G}_0^{[-]}  \right]
		\right)^\rM  \\
& \qquad		
	\rv{+} 	\frac{\mu_0\, e^{\ii \omega_0 t} }{8} \int \frac{d^{D+1}{\rv{\Pi}}}{(2{\rv{\pi}})^{D+1} } \tr
			\left(\partial_{{\rv{\Pi}}^{i}}\hat{Q}_0^{[0]}  \left[\hat{G}_0^{[+]}   \partial_{{\rv{\Pi}}^{[0}}\hat{Q}_0^{[+]}    \hat{G}^{[+]}  \hat{G}^{[-]}  \partial_{{\rv{\Pi}}^{j]}}\hat{Q}_0^{[-]} \hat{G}_0^{[-]}  \right]
			\right)^\rM \\
& \qquad		
	\rv{+} 	\frac{\mu_0\, e^{\ii \omega_0 t} }{8} \int \frac{d^{D+1}{\rv{\Pi}}}{(2{\rv{\pi}})^{D+1} } \tr
			\left(\partial_{{\rv{\Pi}}^{i}}\hat{Q}_0^{[0]}  \left[\hat{G}_0^{[+]}   \partial_{{\rv{\Pi}}^{[0}}\hat{Q}_0^{[+]}   \hat{G}_0^{[+]}  \partial_{{\rv{\Pi}}^{j]}}\hat{Q}_0^{[+]}  \hat{G}^{[+]} \hat{G}^{[-]}  \right]
			\right)^\rM ,
	\label{delta si M}
\end{eqsplit}
where superscript $M$ means that all Green's functions inside the parenthesis are Matsubara Green's functions \Ref{Mats}. Furthermore, we note now that the square parenthesis can be dropped in \Ref{delta si M}, and the trace regain the cyclic property, which was absent for ``lesser" component. This in particular means that in the low temperature limit
\be
	\Delta^{\bf I}\sigma^{ij} = \Delta^{\bf II}\sigma^{ij},
\ee
as they only differ by cyclic transposition of the argument of the trace.

Finally,
\bes
\Delta\sigma^{ij}
	&\equiv \Delta^{\bf I}\sigma^{ij}+\Delta^{\bf II}\sigma^{ij} +{\rm c.c.}
	\\
& = 	\frac{\mu_0\, e^{\ii \omega_0 t} }{4} \int \frac{d^{D+1}{\rv{\Pi}}}{(2{\rv{\pi}})^{D+1} } \tr
		\left(\partial_{{\rv{\Pi}}^{i}}\hat{Q}_0^{[0]}  \left[
			\hat{G}^{[+]}_0   \hat{G}^{[-]}_0   \partial_{{\rv{\Pi}}^{[0}}\hat{Q}_0^{[-]}    \hat{G}_0^{[-]}  \partial_{{\rv{\Pi}}^{j]}}\hat{Q}_0^{[-]} \hat{G}_0^{[-]}  \right]
		\right)^\rM  \\
& \qquad		
	\rv{+} 	\frac{\mu_0\, e^{\ii \omega_0 t} }{4} \int \frac{d^{D+1}{\rv{\Pi}}}{(2{\rv{\pi}})^{D+1} } \tr
			\left(\partial_{{\rv{\Pi}}^{i}}\hat{Q}_0^{[0]}
				\left[\hat{G}_0^{[+]}   \partial_{{\rv{\Pi}}^{[0}}\hat{Q}_0^{[+]}    \hat{G}^{[+]}_0  \hat{G}^{[-]}_0  \partial_{{\rv{\Pi}}^{j]}}\hat{Q}_0^{[-]} \hat{G}_0^{[-]}  \right]
			\right)^\rM \\
& \qquad		
	\rv{+} 	\frac{\mu_0\, e^{\ii \omega_0 t} }{4} \int \frac{d^{D+1}{\rv{\Pi}}}{(2{\rv{\pi}})^{D+1} } \tr
			\left(\partial_{{\rv{\Pi}}^{i}}\hat{Q}_0^{[0]}  \left[
				\hat{G}_0^{[+]}   \partial_{{\rv{\Pi}}^{[0}}\hat{Q}_0^{[+]}   \hat{G}_0^{[+]}  \partial_{{\rv{\Pi}}^{j]}}\hat{Q}_0^{[+]}  \hat{G}^{[+]}_0 \hat{G}^{[-]}_0  \right]
			\right)^\rM
	+{\rm c.c.},
	\label{delta si F}
\end{eqsplit}
Here the integration is over Euclidean space and the Green's  function is the Matsubara one:
$$
G^\rM _0 = (\ii\omega-H(\rv{\vec{\pi}}) )^{-1},
$$
while
$$
Q^\rM _0 = \ii\omega -H(\rv{\vec{\pi}}),
$$
and the $[\pm]$ superscript is defined in \Ref{[pm]}. \rv{In the last two expressions $\vec{\pi} = (\pi^1,..., \pi^D) = (\Pi^1,...,\Pi^D)$ is the $D$ - dimensional vector with spatial components of $\pi$.}

\subsection{ The case of $2D$ Dirac fermions }

\label{Sect2DDirac}

Let us consider as an example the $2+1$ $D$ system consisting of Weyl fermions (they are also referred to as the $2+1$ $D$ Dirac fermions). Each of them corresponds to
$$
	Q = {\bm 1} \omega -v_F (\sigma^1 \pi^1 + \sigma^2 \pi^2 + \sigma^3 m).
$$
Here $m$ is a mass-type parameter, $\si_i$ are Pauli matrices, and ${\bm 1}$ is a unit $2^*2$ matrix. In equilibrium at $T=0$ Hall conductivity is given by
$$
	\sigma_H = \rv{-}\frac{{\cal N}}{2\pi}.
$$
The contribution to ${\cal N}$ of each Weyl fermion is given by \cite{Volovik2003}
$$
	{\cal N}^{(0)} = \frac{1}{2} {\rm sign} \, m.
$$
For bulk real systems these fermions usually come in pairs. Therefore, the total value of $\cal N$ is integer rather than half-integer.

Thus, for the Weyl fermions we take the following modification of the Hamiltonian:
\begin{equation}
	Q = {\bm 1} \omega-v_F (\sigma^1 \pi^1 + \sigma^2 \pi^2 + \sigma^3 m) + \mu_0 \cos\omega_0 t
\end{equation}
Bare (unperturbed) Keldysh matrix $\hat{\bf Q}^< $ has the form
\begin{equation}
\hat{\bf Q}^< =\begin{pmatrix}
	{\bm 1} \omega-v_F (\sigma^1 \pi^1 + \sigma^2 \pi^2 + \sigma^3 m) +\ii\epsilon&-2\ii\epsilon f(\omega)\, {\bm 1}\\0& {\bm 1} \omega-v_F (\sigma^1 \pi^1 + \sigma^2 \pi^2 + \sigma^3 m) -\ii\epsilon
\end{pmatrix},
\end{equation}
while bare Keldysh Green's  function is
\begin{equation}
\hat{\bf G}^< =\begin{pmatrix}
	({\bm 1} \omega-v_F (\sigma^1 \pi^1 + \sigma^2 \pi^2 + \sigma^3 m) +\ii\epsilon)^{-1}
		&-2\pi \ii \delta({\bm 1} \omega-v_F (\sigma^1 \pi^1 + \sigma^2 \pi^2 + \sigma^3 m))f(\omega)\\
	0&({\bm 1} \omega-v_F (\sigma^1 \pi^1 + \sigma^2 \pi^2 + \sigma^3 m) -\ii\epsilon)^{-1}
\end{pmatrix}.
\end{equation}
In the last expression the meaning of the delta function of matrix is that in the basis in which Hamiltonian is diagonal there are delta functions along the diagonal. 

Using explicit expressions for the Green's function and the Hamiltonian in
\Ref{delta si F} we obtain (either by lengthy straightforward algebra, or using Mathematica package) the following correction due to the time dependent chemical potential to the Hall conductivity
\bes
\rv{\Delta\sigma_H}
& = -	{4 \mu_0 v_F^3} \int \frac{d^2 \vec{\pi} d\omega}{(2\pi)^3 }
		\frac{\omega\, m \omega_0^2  \cos(\omega_0 t) }
			{[(\omega-\ii \omega_0/2)^2+v_F^2(\pi^2+m^2)]^2[(\omega+\ii \omega_0/2)^2+v_F^2(\pi^2+m^2)]^2} .
	\label{delta si WF}
\end{eqsplit}
Oddness with respect to transformation $\omega\to-\omega$ gives us the opportunity to conclude that
\be
\Delta\sigma_H =0.
\ee
In order to obtain this result we need that $\omega_0 < m$. Only in this case the integral in the above expression is convergent.

Moreover, we also obtain that any time dependent perturbation of the form
$$
	\hat{H} \to \hat{H}-\mu_0 g(t)
$$
does not lead to modifications of Hall conductivity as long as function $g(t)$ does not contain harmonics with frequencies larger than bare mass parameter $m$.   This conclusion is valid to the first order in $\mu_0$.

\section{ Hall conductivity out of equilibrium in static systems}

\label{SectInvariant}


Let us consider a system without interactions, which in the absence of external electric field is static, though non-equilibrium, i.e. one-particle Hamiltonian $H$ does not depend on time, while distribution function $f(\pi)$ is not of thermal equilibrium. We require, however, that this distribution depends on energy only $f(\pi)\equiv f(\pi^0)$. In such a system we are able to simplify further the expression for the Hall conductivity obtained above:
\begin{equation}
    \sigma_H = \rv{-} \frac{\epsilon^{ij}}{8} \int \frac{d^3\pi}{(2\pi)^3 } \tr \left(\partial_{\pi^{i}}\hat{Q}  \left[\hat{G} \star \partial_{\pi^{[0}}\hat{Q}  \star \partial_{\pi^{j]}}\hat{G}  \right]\right)^< +{\rm c.c.}
\end{equation}
Here the elements of matrices $\hat{Q}$ and $\hat{G}$ are
\bes
Q^{\rR/\rA} (x,\pi) & =  \pi^0-{H}_W(x,\pi)\pm \ii 0
	\\
Q^{<}(x,\pi) &=(Q^\rA (x,\pi) -Q^\rR (x,\pi) )f(\pi^0)
	\\
G^{\rR/\rA} (x,\pi) &=
	\(Q^{\rR/\rA} (x,\pi)\)^{-1}
	\\
G^<(x,\pi)
	&= (G^\rA (x,\pi)-G^\rR (x,\pi)) f(\pi^0)
\end{eqsplit}
Note that the inverse here is understood in respect to the $\star$-product \cite{FZ2019_2}. These functions do not depend on time. At the same time we may impose periodic boundary conditions in spatial coordinates. Therefore, we obtain for the average Hall conductivity
\begin{equation}
\bar{ \sigma}_H
	= \rv{-} \frac{\epsilon^{ij}}{8 \, {\cal V}} \int \frac{d^3\pi d^2 x}{(2\pi)^3 } \tr\left(\partial_{\pi^{i}}\hat{Q}  \star \left[\hat{G} \star \partial_{\pi^{[0}}\hat{Q}  \star \partial_{\pi^{j]}}\hat{G}  \right]\right)^< +{\rm c.c.}
\end{equation}
Here ${\cal V} $ is the overall volume of the system.  The star is inserted here because of the periodic boundary conditions, see \cite{Zubkov+Wu_2019}.  Let us introduce the following notations:
\begin{equation}
  {\cal K}_{\rv{\mu\nu\rho}} =  \frac{1}{4 \, {\cal V}} \int \frac{d^3\pi d^2 x}{(2\pi)^3 } \tr\left(\partial_{\pi^{\mu}}\hat{Q}  \star \hat{G} \star \partial_{\pi^{\nu}}\hat{Q}  \star \partial_{\pi^{\rho}}\hat{G}  \right)^< +{\rm c.c.}\label{K}
\end{equation}
Tensor ${\cal K}^{\mu\nu\rho}$ may be considered as a certain generalization of conductivity.
We have the following expression for electric current density averaged over the area of the system:
$$
\bar{J}^\mu =  { \cal K}^{\mu\nu\rho} {\cal F}_{\nu\rho} = \frac{1}{2} ({ \cal K}^{\mu\nu\rho} -{ \cal K}^{\mu\rho\nu} ) {\cal F}_{\nu\rho}
$$
It contains both response to external electric field and response to external magnetic field. Besides, we may consider the $0$-th component of electric current $J^\mu$ as charge density. That's why the above equation also contains response of electric charge density  to external electromagnetic field.

In Appendix \ref{AppU} we derive expressions for tensor ${\cal K}^{\mu\nu\rho}$ and both Hall and symmetric conductivities:
\begin{equation}
	\bar{\sigma}_H = \rv{-}\frac{{\cal N}_f}{2\pi} + \bar{\sigma}_{H, f^\prime} ,
	\label{MAIN1}
\end{equation}
with
\begin{eqnarray}\label{MAIN2}
    {\cal N}_f &=&
   -\frac{1}{48\pi^2 \, {\cal V}} \epsilon^{\mu\nu\rho} \oint d\pi^0 \int {d^2  \pi d^2 x} \,
   \tr\left(\partial_{\pi^{\mu}}{Q}  \star  \partial_{\pi^{\nu}} {G} \star \partial_{\pi^{\rho}}{Q}  \star G \right) f(\pi^0) +{\rm c.c.},
\end{eqnarray}
where $\oint$ is an integral over the contour surrounding the real axis in the positive direction, while
\bes
	& Q(x,\pi)  = \pi^0-{H}_W(x,\pi), \\
	& Q(x,\pi)\star G(x,\pi) =1.
\end{eqsplit}
At the same time
\begin{eqnarray}
\bar{\sigma}_{H, f^\prime} &=&
	\rv{+}\frac{1}{8 \, {\cal V}}\epsilon^{ij} \int \frac{d^3\pi d^2 x}{(2\pi)^3 } \,
	\tr\left((\partial_{\pi^{i}}Q^\rR \star  {G}^\rR + \partial_{\pi^{i}}Q^\rA \star{G}^\rA ) \star \partial_{\pi^{j}}{Q}^\rR  \star ({G}^\rA -{G}^\rR)  \right) \partial_{\pi^{0}} f(\pi^0)+ c.c.
\end{eqnarray}
is the contribution to Hall conductivity proportional to the derivative of the distribution $f(\pi^0)$. Notice that the similar expression has been obtained in \cite{Mokrousov}.

Expression for ${\cal N}_f$ resembles the topological invariant of zero temperature equilibrium theory. However, the present expression is only partially topological. It remains robust to a limited class of modifications of the system -- to those modifications, that leave initial distribution $f(\pi^0)$ unchanged, and leave one-particle Hamiltonian independent of time. This has been proven in Appendix \ref{AppF}.

For the ordinary symmetric conductivity we obtain
\begin{eqnarray}
    \bar{\sigma}_\parallel^{ij} &=&
            \frac{1}{8 \, {\cal V}} \int \frac{d^3\pi d^2 x}{(2\pi)^3 }
            \tr\left((-\partial_{\pi^{i}}Q^\rR \star  {G}^\rR  + \partial_{\pi^{i}}Q^\rA \star {G}^\rA ) \star \partial_{\pi^{j}}{Q}^\rR  \star ({G}^\rA -{G}^\rR )  \right) \partial_{\pi^{0}} f(\pi^0) + (i \leftrightarrow j) + {\rm c.c.}
\end{eqnarray}
This expression does not contain the term proportional to distribution $f(\pi^0)$, it is expressed through its derivative only.

In Appendix \ref{AppV} we derive expressions for the conductivities through the matrix elements of
velocity operator ${\hat v}_i = \ii [{\cal H}, {\hat x }_i]$. We obtain:
\begin{eqnarray}
{\cal N}_f &=&    -\frac{2 \pi \ii }{4 {\cal V}}\,  \epsilon^{ij} \,
	\sum_{n,k} \Bigl(  \frac{ f^\prime({\cal E}_n)}{({\cal E}_n-{\cal E}_k)} + \frac{f({\cal E}_k)-f({\cal E}_n)}{({\cal E}_k-{\cal E}_n)^2}\Bigr)
		  \langle n|{\hat v}_j| k \rangle    \langle k | {\hat v}_i | n \rangle + {\rm c.c.}
	\label{sigmaHH}
\end{eqnarray}
and
\begin{eqnarray}
\bar{\sigma}_{H,f^\prime} &=&    \frac{\ii \epsilon^{ij} }{4 {\cal V}}\,
	\sum_{n,k} \Bigl( f^\prime({\cal E}_n){\cal P}\frac{1}{({\cal E}_n-{\cal E}_k )}\Bigr)
		  \langle n|{\hat v}_j| k \rangle    \langle k | {\hat v}_i | n \rangle + {\rm c.c.}
	\label{sigmaHH}
\end{eqnarray}
In both expressions, the term with $n=k$ in the discrete spectrum case, must be understand as a limit. In particular, ${\cal P}\frac{1}{({\cal E}_n-{\cal E}_k )}_{k=n} = 0$. Alternatively the total expression for Hall conductivity may be represented as a $\epsilon\to0$ limit of the following expression:
\be
	\si_H =
		-\frac{\ii\, \epsilon^{ij} }{2 \, {\cal V}}\,\sum_{n,k}
		 		\frac{ f({\cal E}_k)}{({\cal E}_k-{\cal E}_n-i\epsilon )({\cal E}_k-{\cal E}_n+i\epsilon)}	
		 		\langle n | \hat v_i | k \rangle
		 		\langle k | \hat v_j | n \rangle
		 		+ {\rm c.c.}
\ee
Being written in this form it coincides with Eq. (1) of \cite{Aoki}.

The following remarks are in order. First of all, expression for ${\cal N}_f$ is ``almost topological", which means that it is robust to smooth modifications of Hamiltonian that do not affect the initial distribution $f$. This has been proven in Appendix \ref{AppF}. Next, an extra contribution to Hall conductivity $\sigma_{H,f^\prime}$ vanishes in thermal equilibrium at $T\to 0$ in the gapped system when Fermi energy is inside the gap. However, this contribution in general case remains nonzero both in equilibrium at $T >0$ and out of equilibrium. As for the symmetric part of the conductivity $\sigma^{ij}_\parallel$ it vanishes as well for the gapped system in equilibrium at $T=0$. At the same time in general case expression for $\sigma^{ij}_\parallel$ may contain an infinity. Its appearance reflects the fact that in the absence of interactions quasiparticles from the zone, which is not occupied completely, are accelerated by external electric field. If we do not take into account interactions (or scattering on impurities) there may exist a divergency in the conductivity.

As for the Hall conductivity, the condition, under which $\sigma_{H, f^\prime}$ may be neglected is that the typical scale of the given system $\Delta$ is small compared to the scale that characterizes distribution $f$. In particular, for thermal equilibrium this condition reads
{$
\Delta \gg T
$.}

In Appendix \ref{AppH} we derive the final expression for the Hall conductivity in the system of non-relativistic electrons in the presence of external magnetic field $\cal B$. In case of thermal equilibrium this expression is simplified further:
\begin{eqnarray}
\bar{\sigma}_H &=& \frac{1}{2\pi } \,\sum_{q \geq 0} \, \frac{1}{e^{{\cal E}_q/T}+1}
.
  \label{sigmaHH}
\end{eqnarray}
Here
$$
{\cal E}_q =  \frac{\cal B}{2m} (2q+1)-\mu
$$
and  $\mu$ is chemical potential.
One can see, that at $T \ll \frac{\cal B}{m}$ this expression is reduced to the zero temperature expression $\bar{\sigma}_H = \frac{N}{2\pi }$, where $N$ is the number of occupied Landau Levels.
Thus, the topological expression remains valid as long as this condition is satisfied, i.e. $T \ll {\cal B}/m$. Eq. (\ref{sigmaHH}) repeats a conventional expression to be read off, for example, from \cite{Kubo}.

In Appendix \ref{SectI} we calculate Hall conductivity for the system of massive $2D$ Dirac fermions (see also Sect. \ref{Sect2DDirac}). The calculation gives for the case of thermal equilibrium with  vanishing Fermi energy:
\begin{eqnarray}
\sigma_H &=& \rv{-}\frac{\alpha}{4\pi} \int_{|\alpha|}^{+\infty} \frac{du}{u^2 \, {\rm th}(u/2)}
\end{eqnarray}
with $\alpha = v_F m/T$. One can see, that this expression tends to $\rv{-}\frac{1}{4\pi}{\rm sign}\,m $ at $T\to 0$. Here the condition for the validity of topological expression for Hall conductivity is $v_F m\gg T$.

\section{Interaction corrections to electric current and Hall conductivity} \label{SectInteraction}

Let us recall expression for electric current in non-interacting systems 
\be
J^i(X)
	\equiv \langle j^i(t,x) \rangle
=
\rv{-}	\frac{\ii }{2}\int \frac{d^{\rv{D+1}}\pi}{(2\pi)^{\rv{D+1}}} \tr\left((\partial_{\pi_{i}}\hat{Q})\hat{G}\right)^{<}
	\rv{-} \frac{\ii }{2}\int \frac{d^{\rv{D+1}}\pi}{(2\pi)^{\rv{D+1}}}
		\tr\left(\hat{G} (\partial_{\pi_{i}}\hat{Q})\right)^{<}.
\ee
Consider now the case, when interactions are present. If, as it is in all reasonable cases, the interaction vertex do not depend on the external electromagnetic potential, the total current is given by
 \begin{equation}\label{J Wigner_0}
J^i(X)
	=
\rv{-}	\frac{\ii }{2}\int \frac{d^{\rv{D+1}}\pi}{(2\pi)^{\rv{D+1}}} \tr\left((\partial_{\pi_{i}}\hat{Q})\hat{\cal G}\right)^{<}
	\rv{-}\frac{\ii }{2}\int \frac{d^{\rv{D+1}}\pi}{(2\pi)^{\rv{D+1}}} \tr\left(\hat{\cal G} (\partial_{\pi_{i}}\hat{Q})\right)^{<}.
\end{equation}
Here $\hat{Q}$ is noninteracting Dirac operator while $\hat{\cal G}$ is the total interacting Green's  function. We shall demonstrate now that it can also be given in terms of interacting Dirac operator $\hat{\cal Q}$:
 \begin{equation}\label{J Wigner_}
\tilde{J}^i(X)
	= \rv{-}\frac{\ii }{2}
		\int \frac{d^{\rv{D+1}}\pi}{(2\pi)^{\rv{D+1}}} \tr\left((\partial_{\pi_{i}}\hat{\cal Q})\hat{\cal G}\right)^{<}
	\rv{-}\frac{\ii }{2}
		\int \frac{d^{\rv{D+1}}\pi}{(2\pi)^{\rv{D+1}}} \tr\left(\hat{\cal G} (\partial_{\pi_{i}}\hat{\cal Q})\right)^{<}.
\end{equation}
The difference between the two is given by
 \begin{equation}\label{delta J}
\Delta {J}^i(X)
	= \rv{-}\frac{\ii }{2}	
		\int \frac{d^{\rv{D+1}}\pi}{(2\pi)^{\rv{D+1}}} \tr\left((\partial_{\pi_{i}}\hat{\Sigma})\hat{\cal G}\right)^{<}
	\rv{-}\frac{\ii }{2}\int \frac{d^{\rv{D+1}}\pi}{(2\pi)^{\rv{D+1}}} \tr\left(\hat{\cal G} (\partial_{\pi_{i}}\hat\Sigma)\right)^{<}.
\end{equation}
Here $\hat\Sigma$ is the self-energy operator, $\hat{\cal Q} = \hat Q - \hat\Sigma$. Let us assume that the interaction has the form of an exchange by a bosonic excitation with homogeneous and $PT$-even propagator $D(p)=D(-p)$, and the simplest possible vertex function equal to unity. Then $\hat\Sigma$ in one loop approximation is given by
$$
	\hat\Sigma(X,\pi) = \int \frac{d^{\rv{D+1}}p}{(2\pi)^{\rv{D+1}}} D(p) \hat{G}(X,\pi-p).
$$
As a result we write \Ref{delta J} integrating by parts and shifting the integration variable
 \begin{eqnarray}\label{J Wigner_}
\Delta {J}^i(X)
&=&
\rv{-}	\frac{\ii }{2}\int \frac{d^{\rv{D+1}}\pi}{(2\pi)^{\rv{D+1}}} \frac{d^{\rv{D+1}}p}{(2\pi)^{\rv{D+1}}} D(p)
		\[\tr\left((\partial_{\pi_{i}} \hat{G}(X,\pi-p) )\hat{G}(X,\pi)\right)^{<}
	\rv{-} \tr\left(\hat{G}(X,\pi) \partial_{\pi_{i}}\hat{G}(X,\pi-p)\right)^{<}
		\]\nonumber\\
&=&
	+ \frac{\ii }{2}\int \frac{d^{\rv{D+1}}\pi}{(2\pi)^{\rv{D+1}}} \frac{d^{\rv{D+1}}p}{(2\pi)^{\rv{D+1}}} D(p)
		\[\tr\left(  \hat{G}(X,\pi-p) \partial_{\pi_{i}}\hat{G}(X,\pi)\right)^{<} +\tr\left((\partial_{\pi_{i}}\hat{G}(X,\pi)) \hat{G}(X,\pi-p)\right)^{<}\]
		\nonumber\\
&=&
	+ \frac{\ii }{2}\int \frac{d^{\rv{D+1}}\pi}{(2\pi)^{\rv{D+1}}} \frac{d^{\rv{D+1}}p}{(2\pi)^{\rv{D+1}}} D(-p)
		\[\tr\left(  \hat{G}(X,\pi)  \partial_{\pi_{i}}\hat{G}(X,\pi-p)\right)^{<}
		+ \tr\left((\partial_{\pi_{i}}\hat{G}(X,\pi-p)) \hat{G}(X,\pi)\right)^{<}
		\]\nonumber\\
&=&
	-\Delta J^i(X).
\end{eqnarray}
We come to the conclusion that $\Delta J^i(X) = 0$, and \Ref{J Wigner_} is valid indeed for interacting systems. In other words, for interacting systems in the expression for the electric current the renormalized velocity, $\partial_{\pi_{i}}\hat{\cal Q}$, may be used instead of the bare one, $\partial_{\pi_{i}}\hat{Q}$.

Now let us consider one-loop correction to conductivity.
Without interactions we have
\be
\hat Q \approx \hat Q^{(0)}-A_\mu \partial_{p_\mu}\hat  Q^{(0)},
	\qquad
\hat G \approx \hat G^{(0)} + {\cal F}^{\mu\nu}\hat {G}_{\mu\nu}^{(1)},
\ee
where according to \Ref{QGK1}
\be
\hat{G}_{\mu\nu}^{(1)} =
\rv{-}	\ii\left(\hat{G}^{(0)}\star \partial_{\pi^{\mu}}\hat{Q}^{(0)} \star\hat{G}^{(0)}\star \partial_{\pi^{\nu}}\hat{Q}^{(0)}\star \hat{G}^{(0)}-(\mu\leftrightarrow \nu)\right)/{2}.
\ee
Therefore, the self-energy operator can be expanded in powers of $\cal F$ in the following way:
\be
	\hat \Sigma =  \hat \Sigma^{(0)}+\hat \Sigma_{\mu\nu}^{(1)}{\cal F}^{\mu\nu},
\ee
where
\be
\hat \Sigma^{(0)} (X,\pi)
	= \int \frac{d^{\rv{D+1}}P}{(2\pi)^{\rv{D+1}}} D(P) \hat{G}^{(0)}(X,\pi-P),
	\qquad
\hat \Sigma_{\mu\nu}^{(1)} (X,\pi)
	= \int \frac{d^{\rv{D+1}}P}{(2\pi)^{\rv{D+1}}} D(P) \hat{G}_{\mu\nu}^{(1)}(X,\pi-P).
\label{Sigma 1}
\ee
We come to the perturbative expansion of interacting $\hat{\cal Q}$ up to the first order in $\mathcal{F}^{\mu\nu}$
\be
	\hat {\cal Q} = \hat Q - \hat \Sigma^{(0)} - \hat \Sigma_{\mu\nu}^{(1)}{\cal F}^{\mu\nu}.
\ee
To find the first correction to interactive $\hat{\cal G}$, we write it as
\be
	\hat {\cal G} = \hat {\cal G}^{(0)}
		+\frac{1}{2}\hat{\cal G}_{\mu\nu}^{(1)}{\cal F}^{\mu\nu},
	\label{QGKa}
\ee
and use Groenewold equation \Ref{Groe-F}
\be
\(\hat{\cal Q}^{(0)} - \hat \Sigma_{\mu\nu}^{(1)}{\cal F}^{\mu\nu}\)
	\star~
		e^{\rv{-}\ii  \mathcal{F}^{\mu\nu}\overleftarrow{\partial}_{\pi^{\mu}}\overrightarrow{\partial}_{\pi^{\nu}}/2}
	\(\hat {\cal G}^{(0)}
			+\frac{1}{2}\hat{\cal G}_{\mu\nu}^{(1)}{\cal F}^{\mu\nu}\) = 1.
	\label{Greo-Inter}
\ee
Here
\be
	\hat {\cal Q}^{(0)} = \hat Q - \hat \Sigma^{(0)},
	\qquad \hat{\cal G}^{(0)} = (\hat Q - \hat \Sigma^{(0)})^{-1} \approx \hat G + \hat G \hat \Sigma^{(0)} \hat G.
\ee
The solution is (to the leading order in interactions), see \Ref{QGK1}:
\begin{eqnarray}
\hat{\cal G}_{\mu\nu}^{(1)}
& = &
	\hat{\cal G}^{(0)} \star  \hat{\Sigma}_{\mu\nu}^{(1)}\star \hat{\cal G}^{(0)}
	\rv{-}\ii\(
		\hat{\cal G}^{(0)}\star \partial_{\pi^{\mu}}\hat{\cal Q}^{(0)} \star{\cal G}^{(0)}\star \partial_{\pi^{\nu}}\hat{\cal Q}^{(0)}\star \hat{\cal G}^{(0)}
			-(\mu\leftrightarrow \nu)\right)/{2}\nonumber\\
&=&
	\hat{G} \star  \hat{\Sigma}_{\mu\nu}^{(1)}\star \hat{G}
	\rv{-}\ii\(
		\hat{\cal G}^{(0)}\star \partial_{\pi^{\mu}}\hat{\cal Q}^{(0)} \star{\cal G}^{(0)}\star \partial_{\pi^{\nu}}\hat{\cal Q}^{(0)}\star \hat{\cal G}^{(0)}
			-(\mu\leftrightarrow \nu)\right)/{2}.
	\label{cal G1}
\end{eqnarray}
The leading order in interactions here is given by:
\begin{eqnarray}
2 \hat{\cal G}_{\mu\nu}^{(1)\, L}(X,\pi)
&=&
	2\hat{G} \star  \hat{\Sigma}_{\mu\nu}^{(1)}\star \hat{G}
	 \rv{-} \ii
		\hat{G} \star  \hat{\Sigma}^{(0)} \star \hat{G} \star \partial_{\pi^{\mu}}\hat{Q} \star{G}\star \partial_{\pi^{\nu}}\hat{Q}\star \hat{G}
	 \rv{+}  \ii
		\hat{G}\star \partial_{\pi^{\mu}}\hat{\Sigma}^{(0)} \star{G}\star \partial_{\pi^{\nu}}\hat{Q}\star \hat{G}
	\nonumber\\&&
	\rv{-}   \ii
		\hat{G}\star \partial_{\pi^{\mu}}\hat{Q} \star \hat{G}\star  \hat{\Sigma}^{(0)} \star \hat{G}\star \partial_{\pi^{\nu}}\hat{Q}\star \hat{G}
	\rv{+}   \ii
		\hat{G}\star \partial_{\pi^{\mu}}\hat{Q} \star{G}\star \partial_{\pi^{\nu}}\hat{\Sigma}^{(0)}\star \hat{G}
	\nonumber\\&&
	\rv{-}   \ii
		\hat{G}\star \partial_{\pi^{\mu}}\hat{Q} \star{G}\star \partial_{\pi^{\nu}}\hat{Q}\star \hat{G} \star  \hat{\Sigma}^{(0)} \star \hat{G}
	-(\mu\leftrightarrow \nu)
	\nonumber\\
&=&
	\ii \int \frac{d^{\rv{D+1}}P}{(2\pi)^{\rv{D+1}}} D(P)  \hat{G} \star   \Biggl[
		-\left(\hat{G} \star \partial_{\pi^{\mu}}\hat{Q}  \star\hat{G} \star \partial_{\pi^{\nu}}\hat{Q} \star \hat{G} \right)_{\pi-P}
	 \rv{-}
		G\Big|_{\pi-P} \star \hat{G} \star \partial_{\pi^{\mu}}\hat{Q} \star{G}\star \partial_{\pi^{\nu}}\hat{Q}
	\nonumber\\&&
\quad	\rv{+}	\partial_{\pi^{\mu}}G\Big|_{\pi-P}  \star{G}\star \partial_{\pi^{\nu}}\hat{Q}
	\rv{-}   \partial_{\pi^{\mu}}\hat{Q} \star \hat{G}\star  G\Big|_{\pi-P}  \star \hat{G}\star
		\partial_{\pi^{\nu}}\hat{Q}
	\nonumber\\&&
\quad	\rv{+}   \partial_{\pi^{\mu}}\hat{Q} \star{G}\star \partial_{\pi^{\nu}}G\Big|_{\pi-P} 
	\rv{-}   \partial_{\pi^{\mu}}\hat{Q} \star{G}\star \partial_{\pi^{\nu}}\hat{Q}\star \hat{G} \star  G\Big|_{\pi-P}
	\Biggr] \star \hat{G}	-(\mu\leftrightarrow \nu)
 \label{cal G1}
\end{eqnarray}

We obtain the following contribution of interactions to electric current averaged over the volume of the system (the part proportional to the field strength). This contribution vanishes in case of thermal equilibrium at $T\to 0$. Moreover, it vanishes also out of equilibrium in static system with discrete spectrum. In this case we may insert the star in the above expressions:
 \begin{eqnarray}
J^{i(1)}_{\cal F}(X) &=&
	\frac{\mathcal{F}^{\mu\nu} }{4 {\cal V}}\int \frac{d^{D} x d^{\rv{D+1}}\pi}{(2\pi)^{\rv{D+1}}}\frac{d^{\rv{D+1}}P}{(2\pi)^{\rv{D+1}}} D(P)  \tr \Biggl[(\partial_{\pi^{i}}\hat{Q}) \star \hat G \star \Bigl(
	 \left(G\star\partial_{\pi^{\mu}}\hat{Q}  \star\hat{G} \star \partial_{\pi^{\nu}}\hat{Q} \star \hat{G} \right)_{\pi-P}
\nonumber\\&&
	\rv{+}
		G\Big|_{\pi-P} \star \hat{G} \star \partial_{\pi^{\mu}}\hat{Q} \star{G}\star \partial_{\pi^{\nu}}\hat{Q}
	 \rv{-}
		\partial_{\pi^{\mu}}G\Big|_{\pi-P}  \star{G}\star \partial_{\pi^{\nu}}\hat{Q}
	\rv{+}
		\partial_{\pi^{\mu}}\hat{Q} \star \hat{G}\star  G\Big|_{\pi-P}  \star \hat{G}\star \partial_{\pi^{\nu}}\hat{Q}
\nonumber\\&&
	\rv{-}
		\partial_{\pi^{\mu}}\hat{Q} \star{G}\star \partial_{\pi^{\nu}}G\Big|_{\pi-P}
	\rv{+}
		\partial_{\pi^{\mu}}\hat{Q} \star{G}\star \partial_{\pi^{\nu}}\hat{Q}\star \hat{G} \star  G\Big|_{\pi-P}
	\Bigr)\star \hat{G} \Biggr]^< + {\rm c.c.}
\end{eqnarray}
The last expression represents a particular case of extension of the diagram technique of \cite{ZZ2019_FeynRule} to nonequilibrium systems. This technique allows to represent the Feynmann diagrams of non-homogeneous theory in terms of Wigner transformed propagators. It may be transferred directly to the Keldysh formalism. The only modification is that all propagators standing in the diagrams are to be taken in their $2\times 2$ matrix Keldysh form.

In the particular case of the above expression we obtain vanishing result if operators under the trace may be reordered in a cyclic way. This occurs when derivative of distribution function may be neglected (see Appendix \ref{App 1a}).


\section{Conclusions}

\label{SectConclusions}

In the present paper we proceed the line of research related to incorporation of Wigner-Weyl calculus to Keldysh technique (see also previous works in this direction \cite{Shitade,Sugimoto,Sugimoto2006,Sugimoto2007,Sugimoto2008} and references therein). Keldysh technique represents actually the complete nonequilibrium quantum field theory, which has conventional QFT and equilibrium quantum statistical physics as its limiting cases. Therefore, the expected applications of Keldysh technique cover all fields of physics, where quantum field theory description is relevant: from condensed matter theory to nuclear physics, elementary particle physics and quantum cosmology. Wigner-Weyl calculus allows to deal with the nonhomogeneous systems as if they are homogeneous. Namely, in \cite{ZZ2019_FeynRule} the diagram technique has been constructed that allows to deal with Wigner transformed propagators instead of the ordinary ones. This technique contains the same amount of integrations over momenta as the one for the homogeneous systems. Using our present approach it may be extended without any modifications to Keldysh QFT. The only change is that the equilibrium propagators are substituted by the Keldysh ones.

Particular cases of application of this technique have been delivered in the present paper. Namely, we perform the gradient expansion for the calculation of fermion propagator and use it to calculate Hall conductivity. This gives Eq. (\ref{MAIN}). It is an extension of the result presented in \cite{Shitade} to non-homogeneous systems.  In \cite{Shitade} its limiting case has been derived for the case when without external electric field the system is homogeneous, although non-equilibrium. Notice that in \cite{Sugimoto2007} all orders of gradient expansion have been discussed. Further we consider the particular case of a non-interacting static non-equilibrium system. In this case the Hall conductivity averaged over the whole volume of the system contains a contribution, which is robust to smooth modification of the system that leaves it static and noninteracting (i.e. the one-particle Hamiltonian in such a system does not depend on time), and does not change the initial distribution. This property may be considered as a ``limited robustness". Actually,  our Eq. (\ref{MAIN2}) is an extension of the previously obtained expression of \cite{Zubkov+Wu_2019} for Hall conductivity of the system that would take place in thermal equilibrium at $T=0$ without external electric field. The non-topological contribution to Hall conductivity may be neglected if derivative of the distribution function may be neglected. We demonstrate that under the same condition the leading order correction to conductivity from an interactions via bosonic excitations vanishes.

Possible extension of our results to the case of non-equilibrium system of general type (when one-particle Hamiltonian depends on time) is a challenge of the work to be done as a continuation of the present study. Working in this direction in the present paper we consider  the system of continuous $2D$ Dirac fermions. In equilibrium at zero temperature each of the fermions  contributes to the Hall conductivity with a term proportional to half of the Klitzing constant. (Notice that in bulk systems there is always an even number of such fermions, so that the total conductivity is proportional to integer multiple of Klitzing constant. However, surface fermions, as in Topological Insulators, can also come in odd numbers.) The dependence of chemical potential on time does not change this property unless there are Fourier harmonics with frequencies larger than the fermion mass. This calculation represents the way how in principle our general expression of Eq. (\ref{MAIN}) may be applied to the consideration of various real processes.

We expect also that the proposed methodology may be useful for the consideration of non-equilibrium counterparts of the other non-dissipative transport phenomena. In particular, the non-equilibrium chiral magnetic effect (CME \cite{Vilenkin,Kharzeev:2009mf,Kharzeev:2009pj,Kharzeev:2013ffa,Kharzeev:2015znc}) may be studied using this technique together with various paths to equilibrium from a non-equilibrium system. Besides, we expect an extension of the approach of \cite{SZ2020} to consideration of chiral separation effect (CSE \cite{Metl,KZ2017}) out of equilibrium. Our technique allows to represent the corresponding conductivities (QHE, CME, CSE) in terms of the Wigner-transformed propagators. In thermal equilibrium at small temperatures all of them are expected to be reduced to the topological expressions. In the present paper we demonstrated this for the case of the QHE. The same refers also to the other members of the family of non-dissipative transport effects \cite{Volovik2003} (say, chiral vortical effect (CVE \cite{Vilenkin,ACVE,Corr1,Corr2}) and chiral torsional effect (CTE \cite{CTE})). In this respect the case of the CME is especially interesting since in true equilibrium its conductivity vanishes \cite{nogo,nogo2}. However, the previous studies indicate that approaching to equilibrium with different orders of limits may restore the naive expression for the CME conductivity in one of the limits (when spatial inhomogeneity is taken off earlier than inhomogeneity in time) \cite{CMEnoneq}. Also it would be interesting to calculate expression for negative magnetoresistance in Dirac/Weyl semimetals using rigorous Keldysh technique. This expression may shed light on the open question about the non-equilibrium version of CME (the one in the presence of parallel electric and magnetic fields \cite{ZrTe5}). The CVE appears to be intimately related to the CSE \cite{ACVE}. Therefore, the technique of \cite{SZ2020} may be transferred to its consideration and adopted also to the Keldysh QFT. CTE represents a certain generalization of the CVE \cite{CTE}. Therefore, our technique may easily be used for its investigation as well. Notice, that the CTE is essentially a finite temperature phenomenon. Therefore, here the calculation of both interaction corrections and corrections due to the non-equilibrium dynamics is especially interesting.

\bibliographystyle{apsrev} 
\bibliography{Ref}

\appendix

\section{Lesser component of a product of operators}

\label{App 1a}
Here we consider the static non-interacting system with the one-particle Hamiltonian that does not depend on time.
We will show that if the derivative of distribution function may be neglected
\be
H \equiv
	\int d^{\rv{D+1}}\pi\, \tr
		\( K_1 \star K_2\star\ldots \star K_n\)^<
	=\int d^{\rv{D+1}}\pi\, f(\pi)
	\[\tr\( K_1 \star K_2\star\ldots \star K_n\)^\rA
		- \tr\( K_1 \star K_2\star\ldots \star K_n\)^\rR \],
\label{KR-KA}
\ee
where each of the $K_i$ is a Green's function $G$, its inverse $Q$ or a derivative of one of those. $f(\pi)$ stands for the initial distribution of the system, not necessarily the equilibrium one. See \Ref{BD<}, \Ref{Gar_expl} and \Ref{Qar_expl} for the notions of $A/R$ and ``lesser" components.

To prove the above, we first note that
\be
	\( K_1 \star K_2\star\ldots K_n\)^<
		= \sum_{i=0}^n \[K_1^\rR \star\ldots\star K_i^\rR  \star K^<_{i+1}
			\star K_{i+2}^\rA \star\ldots\star K_n^\rA \],
\label{Ki<}
\ee
so that in every term of this sum there only one factor in the ``lesser" form.

In the case of a static system for any $K_i$ that does not contain derivatives of $G/Q$, we have (see Eqs. \ref{Gar_expl} and \ref{Qar_expl})
\be
	K^<_i = (K^\rA _i-K^\rR _i) f(\pi).
\ee
Then $H$ becomes
\bes
H &=
	\sum_{i=0}^n  \int d^{\rv{D+1}}\pi\, \tr
		\[K_1^\rR \star\ldots\star K_i^\rR  \star \[(K^\rA _{i+1}-K^\rR _{i+1}) f(\pi)\]
					\star K_{i+2}^\rA \star\ldots\star K_n^\rA \]
		\\
 &=
	\sum_{i=0}^n  \int d^{\rv{D+1}}\pi\, \tr
		\[K_1^\rR \star\ldots\star K_i^\rR  \star \[K^\rA _{i+1} f(\pi)\]
							\star K_{i+2}^\rA \star\ldots\star K_n^\rA \]
		\\
&\qquad\qquad		
		-	\tr\[K_1^\rR \star\ldots\star K_i^\rR  \star \[K^\rR _{i+1} f(\pi)\]
								\star K_{i+2}^\rA \star\ldots\star K_n^\rA \]	.
\end{eqsplit}
This expression can be simplified in the case when position of $f(\pi)$ is irrelevant, i.e. when $K_i\star f(\pi) = K_i f(\pi)$. It is possible in two cases: a) when derivatives of $f(\pi)$ can be assumed being small; b) when all $K_i$ do not depend on time, while $f(\pi)$ depends only on the energy. In the former case, the following expression is only approximate, while in the latter it is exact:
\bes
H &=
	\sum_{i=0}^n  \int d^{\rv{D+1}}\pi\,  f(\pi)\, \tr
		\(K_1^\rR \star\ldots\star K_i^\rR  \star K^\rA _{i+1}
							\star K_{i+2}^\rA \star\ldots\star K_n^\rA  \)\\
&\qquad\qquad
		-	 f(\pi)\, \tr\(K_1^\rR \star\ldots\star K_i^\rR  \star K^\rR _{i+1}
				\star K_{i+2}^\rA \star\ldots\star K_n^\rA \).
\end{eqsplit}
We can see now that in the $i$-th contribution to the sum the first term in the parentheses cancels the second one at $i-1$. Thus, we will be left only with the first and the last terms of the summation:
\bes
H &=
	\int d^{\rv{D+1}}\pi\,  f(\pi)\, \[\tr
		\(K_1^\rA \star\ldots\star K_n^\rA \)
		-	\tr \(K_1^\rR \star\ldots\star K_n^\rR \)\].
\end{eqsplit}

Very similar consideration is valid when
\be
	K_i = \partial_\pi G {\rm\ or\ } K_i = \partial_\pi Q.
\ee
To show it we recall, that due to the mutually inverse nature of $G$ and $Q$, we can always write a derivative of one as a sandwiched derivative of the other,
$$
	\partial_\pi Q = -Q\star \partial_\pi G \star Q,\qquad
		\partial_\pi G = -G\star \partial_\pi Q \star G .
$$
Then, it is sufficient to analyse expressions of the following type entering \Ref{Ki<} in place of one of the $K_i$-s:
\bes
G^\rR  \star \partial_\pi Q^< \star G^\rA
&		= G^\rR  \star \partial_\pi \[(Q^\rA -Q^\rR ) f(\pi)\]\star G^\rA  \\
&		= G^\rR  \star \[\partial_\pi (Q^\rA -Q^\rR )\]  f(\pi)\star G^\rA  +
			G^\rR  \star \[\partial_\pi f(\pi)\](Q^\rA -Q^\rR )\star G^\rA
\end{eqsplit}
Provided that the derivative of $f$ may be neglected, the derivatives of $G$ and $Q$ behave similarly to $G$ ($Q$) themselves in considered construction, and \Ref{KR-KA} holds valid. In Appendix \ref{AppU} we give the more detailed derivation in an important particular case of Hall conductivity, when last term of the above is important.

Finally, we note that disregarding the derivatives of $f(\pi)$ is justified, in particular, when considering the low temperature limit of the initially equilibrium distribution. Moreover, if $f=f(\pi_0)$ is thermal equilibrium distribution indeed, we can further rewrite \Ref{Ki<} as a sum over the Matsubara frequencies.

Indeed, by using the analytic properties of $K^{R/A}$ (as inherited from those of $G/Q$) we have
\be
H	=\int_{{{\rm Im}\pi_0 = \rv{-} 0 }} d^{\rv{D+1}}\pi\, f(\pi_0)
	\tr\( K_1 \star K_2\star\ldots \star K_n\)
		-
		\int_{{{\rm Im}\pi_0 =\rv{+}0 }} d^{\rv{D+1}}\pi\, f(\pi_0)
			\tr\( K_1 \star K_2\star\ldots \star K_n\),
\ee
Closing now the integration contour \rv{over $\pi^0$} into upper (lower) half-plane in the first (second) terms of the above, we transform the original integrals to the two integrals surrounding poles of $f(\pi_0)$. Those are given by imaginary unit times Matsubara frequency $\omega_j = (2j+1)\pi/{\beta}$. Calculating the integral using residue theorem we obtain:
\be
H	=2\pi \rv{i T} \int d^{\rv{D}} \pi \sum_{\om_j}
	\tr\( K_1^\rM  \star K_2^\rM \star\ldots \star K_n^\rM \)
\ee
Here the superscript M indicates that corresponding function is taken at Matsubara frequency: \rv{$K^\rM(\omega_j,\pi_1,...,\pi_D)  \equiv K(i \om_j,\pi_1,...,\pi_D)$}. In the limit of small temperatures the sum over the Matsubara frequencies is reduced to an integral and we arrive at:
\be
H	=\rv{i} \int d^{\rv{D+1}} \rv{\Pi}
	\tr\( K_1^\rM  \star K_2^\rM \star\ldots \star K_n^\rM \).
\ee
\rv{Here $\Pi$ is "Euclidean" $D+1$ - momentum, i.e. $\Pi^{D+1} = \omega$ is continuous Matsubara frequency, $\Pi^i = \pi^i$ for $i=1,...,D$.}

\section{Conductivity in a static system without interactions}

\label{AppU}

Here we derive expressions for Hall and symmetric conductivities through Wigner transformed Green's functions for the case of a  noninteracting static system, in which initial non-thermal distribution depends on energy only in $2+1 D$.

We start from tensor ${\cal K}^{\mu\nu\rho}$ defined in Eq. (\ref{K}):
\begin{equation}
   { \cal K}^{ijk} =  \frac{1}{4 \, {\cal V}} \int \frac{d^3\pi d^2 x}{(2\pi)^3 } \tr \left(\partial_{\pi_{i}}\hat{Q}_0 \star \hat{G}_0\star \partial_{\pi_{\rv{j}}}\hat{Q}_0 \star \partial_{\pi_{\rv{k}}}\hat{G}_0 \right)^< +{\rm c.c.}\label{KU}
\end{equation}

Hall conductivity is given by ${\cal K}^{i0j} - {\cal K}^{ij0}$, so we study both terms separately. We have
\bes
{ \cal K}^{i0j}
&=
	\frac{1}{4 \, {\cal V}} \int \frac{d^3\pi d^2 x}{(2\pi)^3 } \tr \left(\partial_{\pi_{i}}\hat{Q}_0\star \partial_{\pi_{0}}\hat{G}_0 \star \partial_{\pi_{j}}\hat{Q}_0 \star \hat{G}_0 \right)^< +{\rm c.c.}
	\\
&=
    \frac{1}{4 \, {\cal V}} \int \frac{d^3\pi d^2 x}{(2\pi)^3 } \tr \left(\partial_{\pi_{i}} {Q}^\rR _0 \star  \partial_{\pi_{0}} {G}^\rR _0\star \partial_{\pi_{j}} {Q}^\rR _0 \star \Big((G^\rA _0-G^\rR _0) f(\pi_0)\Big) \right)
    \\&
    \quad
    + \frac{1}{4 \, {\cal V}} \int \frac{d^3\pi d^2 x}{(2\pi)^3 } \tr \left(\partial_{\pi_{i}} {Q}^\rR _0 \star  \partial_{\pi_{0}}  {G}^\rR _0\star \partial_{\pi_{j}}\Big((Q^\rA _0-Q^\rR _0) f(\pi_0)\Big) \star  {G}^\rA _0 \right)
    \\&
    \quad
    + \frac{1}{4 \, {\cal V}} \int \frac{d^3\pi d^2 x}{(2\pi)^3 } \tr \left(\partial_{\pi_{i}} {Q}^\rR _0 \star \partial_{\pi_{0}} \Big((G^\rA _0-G^\rR _0) f(\pi_0)\Big)\star \partial_{\pi_{j}} {Q}^\rA _0 \star  {G}^\rA _0 \right)
    \\&
    \quad
    + \frac{1}{4 \, {\cal V}} \int \frac{d^3\pi d^2 x}{(2\pi)^3 } \tr \left(\partial_{\pi_{i}}\Big((Q^\rA _0-Q^\rR _0) f(\pi_0)\Big) \star  \partial_{\pi_{0}}  {G}^\rA _0\star \partial_{\pi_{j}} {Q}^\rA _0 \star  {G}^\rA _0 \right)
    +{\rm c.c.}
    \\
&=
  -\frac{1}{4 \, {\cal V}} \int \frac{d^3\pi d^2 x}{(2\pi)^3 } \tr \left(\partial_{\pi_{i}} {Q}^\rR _0 \star  \partial_{\pi_{0}}  {G}^\rR _0\star \partial_{\pi_{j}} {Q}^\rR _0 \star G^\rR _0 \right) f(\pi_0)
  \\&
  \quad
  + \frac{1}{4 \, {\cal V}} \int \frac{d^3\pi d^2 x}{(2\pi)^3 } \tr \left(\partial_{\pi_{i}}Q^\rA _0 \star  \partial_{\pi_{0}}  {G}^\rA _0\star \partial_{\pi_{j}} {Q}^\rA _0 \star  {G}^\rA _0 \right) f(\pi_0)
  \\&
  \quad
  + \frac{1}{4 \, {\cal V}} \int \frac{d^3\pi d^2 x}{(2\pi)^3 } \tr \left(\partial_{\pi_{i}}Q^\rR _0\star   ( {G}^\rA _0- {G}^\rR _0) \star   \partial_{\pi_{j}} {Q}^\rA _0 \star   {G}^\rA _0 \right) \partial_{\pi_{0}} f(\pi_0)
  +{\rm c.c.}
\label{Ki0j}
\end{eqsplit}
In the last equality we used that position of $f(\pi_0)$ is irrelevant, see App. \ref{App 1a}. Very similarly, for ${\cal K}^{ij0}$ we have
\bes
{ \cal K}^{ij0}
&=
  -\frac{1}{4 \, {\cal V}} \int \frac{d^3\pi d^2 x}{(2\pi)^3 } \tr \left(\partial_{\pi_{i}} {Q}^\rR _0 \star  {G}^\rR _0\star \partial_{\pi_{j}} {Q}^\rR _0 \star \partial_{\pi_{0}}G^\rR _0 \right) f(\pi_0) 
  \\&
  + \frac{1}{4 \, {\cal V}} \int \frac{d^3\pi d^2 x}{(2\pi)^3 } \tr \left(\partial_{\pi_{i}}Q^\rA _0 \star  {G}^\rA _0\star \partial_{\pi_{j}} {Q}^\rA _0 \star \partial_{\pi_{0}} {G}^\rA _0 \right) f(\pi_0)
  \\&
  + \frac{1}{4 \, {\cal V}} \int \frac{d^3\pi d^2 x}{(2\pi)^3 } \tr \left(\partial_{\pi_{i}}Q^\rR _0\star  {G}^\rR _0\star \partial_{\pi_{j}} {Q}^\rR _0 \star ( {G}^\rA _0- {G}^\rR _0)  \right) \partial_{\pi_{0}} f(\pi_0)+{\rm c.c.}
  \label{Kij0}
\end{eqsplit}
Both in \Ref{Ki0j} and \Ref{Kij0} one might be tempted to put in the limit $\epsilon\to0$
$$
	\partial_{\pi_{j}}  {Q}^\rR _0 \star ( {G}^\rA _0- {G}^\rR _0)
	 =
	 	-  {Q}^\rR _0\star \partial_{\pi_{j}}  {G}^\rR _0\star  {Q}^\rR _0 \star ( {G}^\rA _0- {G}^\rR _0)=0,
$$
alleging that $ {Q}^\rR _0 \star ( {G}^\rA _0- {G}^\rR _0) = \cO (\epsilon)$.
While the latter is true, the former is not legitimate since near the pole $\partial_{\pi_{j}}  {G}^\rR _0 \sim 1/\epsilon^2$ and the whole expression is not vanishing.

Thus, for the Hall conductivity we obtain
\bes
\rv{-}\sigma_H
&=
   \frac{1}{2\pi}\times \frac{1}{48\pi^2 \, {\cal V}} \epsilon^{\mu\nu\rho}
  	\int {d^3\pi d^2 x}
  		\tr \left(\partial_{\pi^{\mu}} {Q}^\rR _0 \star  \partial_{\pi^{\nu}}  {G}^\rR _0\star \partial_{\pi^{\rho}} {Q}^\rR _0 \star G^\rR _0 \right) f(\pi^0)
  	\\&
    - \frac{1}{2\pi}\times \frac{1}{48\pi^2 \, {\cal V}} \epsilon^{\mu\nu\rho}\int {d^3\pi d^2 x}
    	\tr \left(\partial_{\pi^{\mu}}Q^\rA _0 \star  \partial_{\pi^{\nu}}  {G}^\rA _0\star \partial_{\pi^{\rho}} {Q}^\rA _0 \star  {G}^\rA _0 \right) f(\pi^0)
    \\&
    + \frac{1}{8 \, {\cal V}} \epsilon^{ij}\int \frac{d^3\pi d^2 x}{(2\pi)^3 }
    	\tr \left(\partial_{\pi^{i}}Q^\rR _0\star   ( {G}^\rA _0- {G}^\rR _0) \star   \partial_{\pi^{j}} {Q}^\rA _0 \star   {G}^\rA _0 \right) \partial_{\pi^{0}} f(\pi^0)
    \\&
    - \frac{1}{8 \, {\cal V}}\epsilon^{ij} \int \frac{d^3\pi d^2 x}{(2\pi)^3 }
    	\tr \left(\partial_{\pi^{i}}Q^\rR _0\star  {G}^\rR _0\star \partial_{\pi^{j}} {Q}^\rR _0 \star ( {G}^\rA _0- {G}^\rR _0)  \right) \partial_{\pi^{0}} f(\pi^0)+{\rm c.c.}
\\&=
	\frac{1}{2\pi} {\cal N}_f
    	+
    	\epsilon^{ij} \cA_{ij}.
\label{sigma_H_f}
\end{eqsplit}
Here
\be
	\cA_{ij} = \frac{1}{8 \, {\cal V}}
		\int \frac{d^3\pi\, d^2 x}{(2\pi)^3 }
		\tr \left(
			\(
				\partial_{\pi^{\rv{j}}}Q^\rA _0\star {G}^\rA _0 \star \partial_{\pi^{\rv{i}}} {Q}^\rR _0
				-\partial_{\pi^{\rv{i}}}Q^\rR _0\star   {G}^\rR _0 \star\partial_{\pi^{\rv{j}}} {Q}^\rR _0
			\)
			  \star ( {G}^\rA _0- {G}^\rR _0)
		\right)
		\partial_{\pi_{0}} f(\pi_0)
	 +{\rm c.c.},
	 \label{cA}
\ee
and
\begin{eqnarray}
    {\cal N}_f &=&
   -\frac{1}{48\pi^2 \, {\cal V}} \epsilon^{\rv{ijk}} \oint d\pi_0 \int {d^2  \vec{\pi} d^2 x} \tr \left( G_0\star  \partial_{\pi^{\rv{i}}} {Q}_0 \star  \partial_{\pi^{\rv{j}}}  {G}_0\star \partial_{\pi^{\rv{k}}} {Q}_0 \right) f(\pi_0) +{\rm c.c.},
\label{cN_f}
\end{eqnarray}
where $\oint$ is an integral over the contour encompassing the whole real axis in positive direction, while
\bes
	Q_0(x,\pi)  &= \pi_0- {H}_W(x,\pi),
	\\
	G_0(x,\pi) &\star   Q_0(x,\pi) =1.
\end{eqsplit}

In a similar way for the longitudinal conductivity we obtain
\be
\sigma_\parallel^{ij} =
     \cA_{\{ij\}}
    + {\rm c.c.}
\ee

\section{Conductivity in terms of the velocity operator}
\label{AppV}

Let us now rewrite \Ref{sigma_H_f} in terms of the matrix elements of the velocity operator,  similar to the derivation given in \cite{Zubkov+Wu_2019}. For this end we shall use that the trace of the Weyl symbols over the phase space is equal to the functional trace of a product of corresponding operators, for instance, given by the trace of their matrix elements over momentum space
\bes
\Tr(A_W \ast B_W)  & \equiv \int d^3 X  \int \frac{d^3 P}{(2\pi)^3}\tr (A_W \ast B_W)
	\\
	&= {\Tr} \hat{A} \hat{B} = \int d^3 P d^3 Q A(P,Q) B(Q,P),
\label{tr AB}
\end{eqsplit}
where \rv{$P^i = (p_0,p_1,p_2) = (\omega,p)$ and $X^i = (t,x_1,x_2) = (t,x)$}. Applying this formula to Eq. \Ref{cN_f} we come to
\bes
\cN_f
 = 	-\frac{\epsilon^{\mu\nu\rho}}{48\pi^2 \, {\cal V}}  \oint d \om^{(1)} \, f(\om^{(1)})
  \int  d^2 p^{(1)} \prod_{i=2}^4 d^3 P^{(i)}
  &\tr \Bigg[
 	\rv{G}(P^{(1)},P^{(2)})
	\[\partial_{P^{(2)}_\mu} + \partial_{P^{(3)}_\mu}\] \rv{Q}(P^{(2)},P^{(3)})
	\\
  	&
   	\( \[\partial_{P^{(3)}_\nu} + \partial_{P^{(4)}_\nu}\]  \rv{G}(P^{(3)},P^{(4)}) \)
 		\[\partial_{P^{(4)}_\rho} + \partial_{P^{(1)}_\rho}\] \rv{Q}(P^{(4)},P^{(1)})
 		\Bigg]
 		+{\rm c.c.}
\end{eqsplit}
For the non-interacting fermions described by Hamiltonian ${\cal H}$ with energy eigenstates $|n\rangle$: $\cH |n\rangle = \cE_n |n\rangle$,  the matrix elements in the above are given by
\be
Q(P^{(1)},P^{(2)}) \equiv \langle P^{(1)}| \hat{Q} | P^{(2)}\rangle
	= \(
		\delta^{(2)} (p^{(1)}-p^{(2)})  \omega^{(1)}
		- \langle p^{(1)}| {\cal H} | p^{(2)}\rangle
		 \) \delta(\omega^{(1)}-\omega^{(2)})
\label{QG}
\ee
$$
G(P^{(1)},P^{(2)}) =
	\delta(\omega^{(1)}-\omega^{(2)})
	\sum_{n} \frac{\langle p ^{(1)}| n \rangle \langle n | p ^{(2)}\rangle}{\omega^{(1)}-{\cal E}_n} .
$$
Here $\sum_n$ may stand both for discrete spectrum summation, and integration $\int dn$ in the case of continuum one.

To perform further simplifications, we note that
\be
	\partial_{p_i} G = - G \partial_{p_i} Q G ,
\ee
and more importantly,
$$
	\[\partial_{p^{(4)}_j} + \partial_{p^{(1)}_j} \]
		\langle p ^{(4)}| {\cal H} | p ^{(1)} \rangle
	= \ii   \langle p ^{(4)}| {\cal H} {\hat x }_j -{\hat x }_j{\cal H}   | p ^{(1)}\rangle
	\equiv
	\langle p ^{(4)}| \hat v_j | p ^{(1)}\rangle,
$$
where we introduced the velocity operator ${\hat v}_i = \ii [{\cal H}, {\hat x }_i]$.
By operator $\hat x_i$ we understand $\ii\partial_{p_i}$ acting on the wavefunction written in momentum representation:
$$
\hat{x}_j \Psi(p)
	= \langle p|\hat{x}_j |\Psi\rangle
	= \ii\partial_{p_j} \langle p|\Psi\rangle
	= \ii \partial_{p_j} \Psi(p).
$$
Then, for~example,
$$
\hat{x}_j \delta^{(2)}(q-p)
	= \langle p|\hat{x}_j |q\rangle
	= \ii\partial_{p_j} \langle p|q \rangle
	= \ii \partial_{p_j} \delta^{(2)}(p-q)
	= -\ii \partial_{p_j}\langle q|p \rangle.
$$
Therefore, we can write
$$
\hat{x}_j |p\rangle = -\ii\partial_{p_j} |p \rangle.
$$

Using the above formulae, we derive that
\bes
\cN_f &
 = \rv{+}\frac{\epsilon^{ij} }{4\, {\cal V}}\,\sum_{n,k} \int  \prod_{l=1}^4 d^2 p ^{(l)}
 \oint \frac{ f(\omega) d \omega }{(\omega^{}-{\cal E}_n)^2 (\omega^{}-{\cal E}_k)}
 	\langle p ^{(1)}| n \rangle \langle n p ^{(2)}\rangle
 		\langle p ^{(2)}| \hat v_i | p ^{(3)}\rangle
 		\langle p ^{(3)}| k \rangle \langle k | p ^{(4)}\rangle
 		\langle p ^{(4)}| \hat v_j | p ^{(1)}\rangle
 + {\rm c.c.}
 \\
 &
 = \rv{+}\frac{2\pi \ii\, \epsilon^{ij} }{4 \, {\cal V}}
 		\sum_{n,k }
 		\frac{ f({\cal E}_k)-f({\cal E}_n) + ({\cal E}_n-{\cal E}_k) f'({\cal E}_n)}{({\cal E}_k-{\cal E}_n)^2} 
 		\langle n | \hat v_i | k \rangle
 		\langle k | \hat v_j | n \rangle
 		+ {\rm c.c.}
\end{eqsplit}
We used here that the momentum eigenvalues compose a full set, $\int d^2p\, |p\rangle \langle p|=1$. Note, that in the case of a discreet spectrum, the term $n=k$ should be understood as a limit ${\cal E}_n\to{\cal E}_k$, which gives a finite result.

For the non-topological contribution to $\si_H$ and for $\si_\|$ we shall similarly analize $\cA$ given by \Ref{cA}.
Advanced and retarded components needed for its calculation  can be obtained from \Ref{QG} as
\be
Q^{\rA/\rR}(p^{(1)},p^{(2)}) 
		= \( \delta^{(2)} (p^{(1)}-p^{(2)})\,  \[\omega^{(1)}\pm \ii\epsilon \]
			- \langle p^{(1)}| {\cal H} | p^{(2)}\rangle \) \delta(\omega^{(1)}-\omega^{(2)})
\ee
$$
	G^{\rA/\rR}(P^{(1)},P^{(2)}) = \delta(\omega^{(1)}-\omega^{(2)})
		\sum_{n} \frac{\langle p ^{(1)}| n \rangle \langle n | p ^{(2)}\rangle}{\omega^{(1)}-{\cal E}_n\pm \ii\epsilon} .
$$
and thus,
\be
	({G}^\rA _0-{G}^\rR _0)(P^{(1)},P^{(2)})
		=2\pi \ii\, \delta(\omega^{(1)}-\omega^{(2)}) \sum_{n} \delta(\omega^{(1)}-{\cal E}_n)  \langle p ^{(1)} | n \rangle \langle n | p ^{(2)}\rangle .
\ee
Then
\be
(\partial_{\pi_{j}}Q^\rA _0 \hat{G}^\rA _0  \partial_{\pi_{i}}\hat{Q}^\rR _0 )(P^{(1)},P^{(4)})
	=
	\delta(\omega^{(1)}-\omega^{(4)})
	\int dp^{(2)}dp^{(3)}
	\langle p ^{(1)}| \hat v_j | p ^{(2)}\rangle
	\frac{\langle p ^{(2)}| n \rangle \langle n | p^{(3)}\rangle}{\omega^{(1)}-{\cal E}_n + \ii\epsilon}
	\langle p ^{(3)}| \hat v_i | p ^{(4)}\rangle ,
\ee
and
\be
(\partial_{\pi_{i}}Q^\rR _0 \hat{G}^\rR _0  \partial_{\pi_{j}}\hat{Q}^\rR _0 )(P^{(1)},P^{(4)})
	=
	\delta(\omega^{(1)}-\omega^{(4)})
	\int dp^{(2)}dp^{(3)}
	\langle p ^{(1)}| \hat v_i | p ^{(2)}\rangle
	\frac{\langle p ^{(2)}| n \rangle \langle n | p^{(3)}\rangle}{\omega^{(1)}-{\cal E}_n - \ii\epsilon}
	\langle p ^{(3)}| \hat v_j | p ^{(4)}\rangle .
\ee
All together it gives
\be
	\cA_{ij} = \rv{-} \frac{ \ii }{8 \, {\cal V}}
	\sum_{n,k} f' ({\cal E}_k)
		\[\frac{\langle  k| \hat v_j | n \rangle \langle n | \hat v_i | k \rangle}{{\cal E}_k-{\cal E}_n + \ii\epsilon}
		-\frac{\langle  k| \hat v_i | n \rangle \langle n | \hat v_j | k \rangle}{{\cal E}_k-{\cal E}_n - \ii\epsilon}\]
	 +{\rm c.c.},
\ee
So, that
\bes
\cA_{\{ij\}}
&
	=  \rv{-} \frac{\ii }{8 \, {\cal V}}
	\sum_{n,k} f' ({\cal E}_k)
		\[\frac{1}{{\cal E}_k-{\cal E}_n + \ii\epsilon}
		-\frac{1}{{\cal E}_k-{\cal E}_n - \ii\epsilon}\]
		\(\langle  k| \hat v_j | n \rangle \langle n | \hat v_i | k \rangle+\langle  k| \hat v_i | n \rangle \langle n | \hat v_j | k \rangle\)
	 +{\rm c.c.},
	 \\
\cA_{[ij]} &	
	= \rv{-} \frac{ \ii }{8 \, {\cal V}}
		\sum_{n,k} f' ({\cal E}_k)
			\[\frac{1}{{\cal E}_k-{\cal E}_n + \ii\epsilon}
			+ \frac{1}{{\cal E}_k-{\cal E}_n - \ii\epsilon}\]
			\(\langle  k| \hat v_j | n \rangle \langle n | \hat v_i | k \rangle
				- \langle  k| \hat v_i | n \rangle \langle n | \hat v_j | k \rangle\)
		 +{\rm c.c.},
\end{eqsplit}
The limit $\epsilon\to0$ of these expressions depends on the nature of the spectrum. In the continuum case, the Sokhotski-Plemelj formula gives
\be
	\frac{1}{{\cal E}_k-{\cal E}_n + \ii\epsilon}
		- \frac{1}{{\cal E}_k-{\cal E}_n - \ii\epsilon}
			= - 2 \pi \ii \, \delta({\cal E}_k-{\cal E}_n),
	\qquad
	\frac{1}{{\cal E}_k-{\cal E}_n + \ii\epsilon}
		+ \frac{1}{{\cal E}_k-{\cal E}_n - \ii\epsilon}
			= 2 {\cal P}\frac{1}{{\cal E}_k-{\cal E}_n},
\ee
while in the discrete case, the expression for $\cA_{\{ij\}}$ (and thus, for symmetric conductivity) will be divergent in $\epsilon \to0$
\be
\frac{1}{{\cal E}_k-{\cal E}_n + \ii\epsilon}
	- \frac{1}{{\cal E}_k-{\cal E}_n - \ii\epsilon}
	=
	\left\{\begin{array}{ll}
	0, & n\ne k\\
	\frac2{\ii \epsilon}, & n =  k
	\end{array}\right., \qquad
\frac{1}{{\cal E}_k-{\cal E}_n + \ii\epsilon}
	+ \frac{1}{{\cal E}_k-{\cal E}_n - \ii\epsilon}
	=
	\left\{\begin{array}{ll}
	\frac{2}{{\cal E}_k-{\cal E}_n}, & n\ne k\\
	0, & n =  k
	\end{array}\right.	
\ee

Summarizing, we have
\be
	\si_H = 
		\rv{- \sigma_{xy} = } - \frac{\ii\, \epsilon^{ij} }{4 \, {\cal V}}\,\sum_{n,k}
		 		\(\frac{ f({\cal E}_k)-f({\cal E}_n) + ({\cal E}_n-{\cal E}_k) f'({\cal E}_n)}{({\cal E}_k-{\cal E}_n)^2}
		 			+   f'({\cal E}_k) {\cal P}\frac1{{\cal E}_k-{\cal E}_n}\)
		 		\langle n | \hat v_i | k \rangle
		 		\langle k | \hat v_j | n \rangle
		 		+ {\rm c.c.}\label{C14}
\ee
We can use this expression both for continuum and discreet spectrum if in the latter case we put ${\cal P}\frac1{{\cal E}_k-{\cal E}_n}=0$, ${\cal E}_n={\cal E}_k$. One can see, that in the absence of the singularities at ${\cal E}_n = {\cal E}_k$ the term with $f^\prime$ is cancelled. In Eq. (\ref{C14}) the singularity is isolated in the second term in the brackets while the first term remains finite at ${\cal E}_n = {\cal E}_k$ (it is reduced to $ f^{\prime \prime}({\cal E}_n)/2$). It is worth mentioning that one can rewrite the whole expression in the following alternative form:
\be
	\si_H =
		\rv{-}\frac{\ii\, \epsilon^{ij} }{2 \, {\cal V}}\,\sum_{n,k}
		 		\frac{ f({\cal E}_k)}{({\cal E}_k-{\cal E}_n-i\epsilon )({\cal E}_k-{\cal E}_n+i\epsilon)}	
		 		\langle n | \hat v_i | k \rangle
		 		\langle k | \hat v_j | n \rangle
		 		+ {\rm c.c.}\label{C15}
\ee
Written in this form it coincides with expression proposed in \cite{Aoki} (see also \cite{KuboSigma}). In order to show equivalence of Eqs. (\ref{C15}) and (\ref{C14}) let us represent the quotient from the former as follows:
\bes
&  \frac{ f({\cal E}_k)-f({\cal E}_n)}{({\cal E}_k-{\cal E}_n-i\epsilon )({\cal E}_k-{\cal E}_n+i\epsilon)}	
	\\
&\qquad =
	\frac{  f({\cal E}_k)-f({\cal E}_n) + ({\cal E}_n-{\cal E}_k) f^{\prime}({\cal E}_n)}{({\cal E}_k-{\cal E}_n-i\epsilon )({\cal E}_k-{\cal E}_n+i\epsilon)} - \frac{ ({\cal E}_n-{\cal E}_k -i\epsilon) f^{\prime}({\cal E}_n)}{2({\cal E}_k-{\cal E}_n-i\epsilon )({\cal E}_k-{\cal E}_n+i\epsilon)}-\frac{   ({\cal E}_n-{\cal E}_k+i\epsilon) f^{\prime}({\cal E}_n)}{2({\cal E}_k-{\cal E}_n-i\epsilon )({\cal E}_k-{\cal E}_n+i\epsilon)}
	\\
&\qquad =
	\frac{  f({\cal E}_k)-f({\cal E}_n) + ({\cal E}_n-{\cal E}_k) f^{\prime}({\cal E}_n)}{({\cal E}_k-{\cal E}_n-i\epsilon )({\cal E}_k-{\cal E}_n+i\epsilon)} + \frac{  f^{\prime}({\cal E}_n)}{2({\cal E}_k-{\cal E}_n+i\epsilon)}+\frac{   f'({\cal E}_n)}{2({\cal E}_k-{\cal E}_n-i\epsilon )}	
	\\
&\qquad =
	\frac{  f({\cal E}_k)-f({\cal E}_n) + ({\cal E}_n-{\cal E}_k) f^{\prime}({\cal E}_n)}{({\cal E}_k-{\cal E}_n-i\epsilon )({\cal E}_k-{\cal E}_n+i\epsilon)}
		+ f'({\cal E}_n) {\cal P}\frac{1 }{({\cal E}_k-{\cal E}_n)}
\label{C16}
\end{eqsplit}

\section{Hall conductivity for the noninteracting 2D system in the presence of constant magnetic field}
\label{AppH}

Here we demonstrate how the derived expressions allow to obtain final expressions for the conductivity. We take as an example the simplest system of free non-relativistic electrons in the presence of constant magnetic field. The one-particle Hamiltonian is taken in its simplest form
$$
{\cal H}  = \frac{1}{2m}(\pi_1^2 + \pi_2^2)-\mu
$$
with $\pi_1 = \hat{p}_1$ and $\pi_2 =\hat{p}_2-{\cal B}x_1$. We have the following property specific for this Hamiltonian to be used further:
$$
\epsilon^{ij}\pi_i {\cal H} \pi_j = 3i{\cal B} {\cal H}
$$
The average Hall conductivity may be represented as
\begin{eqnarray}
\bar{\sigma}_H &=&    -\frac{\ii }{4 {\cal V}}\,  \epsilon^{ij} \,
	   \Big(\sum_{n,k|{\cal E}_n\ne {\cal E}_k}
 		\frac{ f({\cal E}_k)-f({\cal E}_n) }{({\cal E}_k-{\cal E}_n)^2} + \frac{1}{2}\sum_{n,k|{\cal E}_n={\cal E}_k}f^{\prime \prime}({\cal E}_n)\Big)
		  \langle n|{\hat v}_i| k \rangle    \langle k | {\hat v}_j | n \rangle + {\rm c.c.}
	\label{sigmaHH_}
\end{eqnarray}
In order to calculate the value of $\bar{\sigma}_{H}$ we decompose the coordinates $x_1, x_2 $ as follows:
$$
\hat{x}_1 = -\frac{\hat{p}_2-{\cal B}x_1}{\cal B} + \hat{\cal X}_1 = \hat{\xi}_1 + \hat{\cal X}_1,$$ $$ \hat{x}_2 = \frac{\hat{p}_1}{\cal B} + \hat{\cal X}_2= \hat{\xi}_2 + \hat{\cal X}_2\,.
$$
The commutation relations follow:
$$
[\hat{\xi}_1,\hat{\xi}_2] = \frac{i}{\cal B}\,, \quad [\hat{\cal X}_1,\hat{\cal X}_2] =-\frac{i}{\cal B}\,,
$$
$$
[{\cal H}, \xi_1] = -i  \frac{\partial}{\partial p_1} {\cal H}\,, \quad [{\cal H}, \xi_2] =  -i \frac{\partial}{\partial p_2} {\cal H}\,,
$$
$$
[{\cal H}, \hat{\cal X}_1] =  [{\cal H}, \hat{\cal X}_2] =  0\,.
$$
Here we use that the Hamiltonian contains the following dependence on $x$:
$$
{\cal H}(\hat{p}_1, \hat{p}_2-{\cal B} x_1)\,
$$
and $\frac{\partial^2}{\partial p_1 \partial p_2}{\cal H}=0$.
Thus we obtain:
\rev{\begin{eqnarray}
\bar{\sigma}_H &=&  -\frac{\ii }{2 {\cal V}}\,  \epsilon^{ij} \,
	 \Big(\sum_{n,k|{\cal E}_n\ne {\cal E}_k}
 		\frac{ f({\cal E}_k)-f({\cal E}_n) + ({\cal E}_n-{\cal E}_k) f'({\cal E}_n)}{({\cal E}_k-{\cal E}_n)^2} + \frac{1}{2}\sum_{n,k|{\cal E}_n={\cal E}_k}f^{\prime \prime}({\cal E}_n)\Big)\langle n| [{\cal H}, {\hat \xi}_i] | k \rangle    \langle k | [{\cal H}, {\hat \xi}_j] | n \rangle
\nonumber\\
&=&  \frac{\ii }{2 {\cal V}}\,  \epsilon^{ij} \,
	  \Bigl(- \sum_{n,k|{\cal E}_n\ne {\cal E}_k} 2f({\cal E}_n)+ \frac{1}{2}\sum_{n,k|{\cal E}_n={\cal E}_k}f^{\prime \prime}({\cal E}_n)({\cal E}_k-{\cal E}_n)^2  \Bigr)\,\Big[  \langle n|  {\hat \xi}_i | k \rangle    \langle k |  {\hat \xi}_j | n \rangle  \Big]
\nonumber\\
&=&  \frac{\ii }{2 {\cal V}}\,  \epsilon^{ij} \,
	   \Bigl(-\sum_{n,k} 2f({\cal E}_n) +  \sum_{n,k|{\cal E}_n={\cal E}_k} 2f({\cal E}_n)\Bigr) \Big[  \langle n|  {\hat \xi}_i | k \rangle    \langle k |  {\hat \xi}_j | n \rangle  \Big]
\nonumber\\
&=&  \frac{\ii }{2 {\cal V}}\,
	  \sum_{n} \Bigl( -2f({\cal E}_n)\Bigr)\,\Big[  \langle n|  [{\hat \xi}_1,  {\hat \xi}_2] | n \rangle  \Big]
\nonumber\\
&=&  \frac{1 }{{\cal B} {\cal V}}\,
	  \sum_{n} f({\cal E}_n)\, \langle n|  n \rangle  = \frac{\rho}{\cal B}.\label{QHEB}
\end{eqnarray}}
Here $\rho$ is density of electrons. We used here that momentum $p_2$ is a good quantum number, and it enumerates the eigenstates of the Hamiltonian:
$$
{\cal H} |n\rangle = {\cal H}(\hat{p}_1, p_y-{\cal B} x_1)|p_2, q\rangle = {\cal E}_{q}|p_2, q\rangle ,
	\quad q\in \mathds{N}\,.
$$
We assume that the size of the system is $L\times L$. Properties of the eigenstates of Hamiltonian guarantee that  $\langle p^\prime_2, q| \hat{p}_1|p_2, q\rangle=0$ for $p^\prime_2 \ne p_2$.
This gives
\rev{\begin{eqnarray}
\bar{\sigma}_H &=& \,\sum_{q}\int \frac{dp_2 L}{2\pi {\cal V}}  \, \frac{ f({\cal E}_q)}{\cal B}.
  \label{calM2d232}
\end{eqnarray}}
$\langle x_1 \rangle = p_2/{\cal B}$ plays the role of the center of orbit, and this center should belong to the interval $(-L/2, L/2)$ while ${\cal E}_q$ does not depend on momentum. This gives
\rev{\begin{eqnarray}
\bar{\sigma}_H &=& \frac{1}{2\pi } \,\sum_{q\in \mathds{N}} \,  f({\cal E}_q).
  \label{calM2d233}
\end{eqnarray}}
In case of thermal equilibrium this expression receives the form:
\begin{eqnarray}
\bar{\sigma}_H &=& \frac{1}{2\pi } \,\sum_{q=0,1,...} \,  \frac{1}{e^{{\cal E}_q/T}+1}.
  \label{calM2d233}
\end{eqnarray}
Here
$$
{\cal E}_q =  \frac{\cal B}{2m} (2q+1)-\mu
$$
where $\mu$ is chemical potential.
One can see, that at $T \ll \frac{\cal B}{m}$ this expression is reduced to the zero temperature expression $\bar{\sigma}_H = \frac{N}{2\pi }$, where $N$ is the number of occupied Landau Levels.

\section{Robustness of ${\cal N}_f$ with respect to modification of one-particle Hamiltonian}

\label{AppF}
Let us consider the following quantity for the 2+1 dimensional system
\begin{eqnarray}
    {\cal N} &=&
    \Big( \frac{\epsilon_{ijk}}{48 \pi^2 \, {\cal V}} \int {d^3\pi d^2 x} \, \tr \left(\hat{G}_0 \star \partial_{\pi_{i}}\hat{Q}_0 \star \hat{G}_0\star \partial_{\pi_{j}}\hat{Q}_0 \star \hat{G}_0\star \partial_{\pi_{k}}\hat{Q}_0  \right)^<  +{\rm c.c.}\Big)\label{MAIN2A}
\end{eqnarray}
We consider the static system with distribution function that depends only  on $\pi_0$. Variation of $\cal N$ caused by a variation of $\hat{Q}$ (such that $\delta f(\pi_0)=0$) gives:
\begin{eqnarray}
  \rv{\delta}  {\cal N} &=&
    \Big( \frac{3\epsilon_{ijk}}{48 \pi^2 \, {\cal V}} \int {d^3\pi d^2 x} \, \tr \left(\delta\hat{G}_0 \star \partial_{\pi_{i}}\hat{Q}_0 \star \hat{G}_0\star \partial_{\pi_{j}}\hat{Q}_0 \star \hat{G}_0\star \partial_{\pi_{k}}\hat{Q}_0  \right)^<  +{\rm c.c.}\Big)\nonumber\\&&+
     \Big( \frac{3\epsilon_{ijk}}{48 \pi^2 \, {\cal V}} \int {d^3\pi d^2 x} \, \tr \left(\hat{G}_0 \star \partial_{\pi_{i}}\delta \hat{Q}_0 \star \hat{G}_0\star \partial_{\pi_{j}}\hat{Q}_0 \star \hat{G}_0\star \partial_{\pi_{k}}\hat{Q}_0  \right)^<  +{\rm c.c.}\Big)\nonumber\\&=&
    \Big( -\frac{3\epsilon_{ijk}}{48 \pi^2 \, {\cal V}} \int {d^3\pi d^2 x} \, \tr \left(\delta\hat{G}_0 \star \partial_{\pi_{i}}\hat{Q}_0 \star \partial_{\pi_{j}}\hat{G}_0 \star  \partial_{\pi_{k}}\hat{Q}_0  \right)^<  +{\rm c.c.}\Big)\nonumber\\&&+
     \Big( -\frac{3\epsilon_{ijk}}{48 \pi^2 \, {\cal V}} \int {d^3\pi d^2 x} \, \tr \left(\hat{G}_0 \star \partial_{\pi_{i}}\delta \hat{Q}_0 \star  \partial_{\pi_{j}}\hat{G}_0 \star \partial_{\pi_{k}}\hat{Q}_0  \right)^<  +{\rm c.c.}\Big)\nonumber\\&=&
    \Big(- \frac{3\epsilon_{ijk}}{48 \pi^2 \, {\cal V}} \int {d^3\pi d^2 x} \, \tr \left(\hat{G}_0 \star \delta\hat{Q}_0 \star \hat{G}_0 \star  \partial_{\pi_{i}}\hat{Q}_0 \star \hat{G}_0 \star  \partial_{\pi_{j}}\hat{Q}_0 \star \hat{G}_0 \star   \partial_{\pi_{k}}\hat{Q}_0  \right)^<  +{\rm c.c.}\Big)\nonumber\\&&+
     \Big( \frac{3\epsilon_{ijk}}{48 \pi^2 \, {\cal V}} \int {d^3\pi d^2 x} \, \tr \left(\hat{G}_0 \star \partial_{\pi_{i}}\hat{Q}_0 \star \hat{G}_0 \star  \delta \hat{Q}_0 \star \hat{G}_0 \star  \partial_{\pi_{j}}\hat{Q}_0 \star \hat{G}_0 \star  \partial_{\pi_{k}}\hat{Q}_0  \right)^<  +{\rm c.c.}\Big)
\end{eqnarray}
Here similar to the case of quantities considered in Appendix \ref{App 1a} and Appendix \ref{AppU} the above expressions obey the cyclic property {\it provided that function $f(\pi_0)$ remains unchanged and the contribution proportional to its derivative may be neglected}. Further we simplify these expressions and obtain:
\begin{eqnarray}
   \rv{\delta} {\cal N} &=&
    \Big(- \frac{3\epsilon_{ijk}}{48 \pi^2 \, {\cal V}} \int {d^3\pi d^2 x} \, \tr \left(\hat{G}_0 \star \delta\hat{Q}_0 \star \hat{G}_0 \star  \partial_{\pi_{i}}\hat{Q}_0 \star \hat{G}_0 \star  \partial_{\pi_{j}}\hat{Q}_0 \star \hat{G}_0 \star   \partial_{\pi_{k}}\hat{Q}_0  \right)^<  +{\rm c.c.}\Big)\nonumber\\&&+
     \Big( \frac{3\epsilon_{ijk}}{48 \pi^2 \, {\cal V}} \int {d^3\pi d^2 x} \, \tr \left( \hat{G}_0 \star  \delta \hat{Q}_0 \star \hat{G}_0 \star  \partial_{\pi_{j}}\hat{Q}_0 \star \hat{G}_0 \star  \partial_{\pi_{k}}\hat{Q}_0 \star \hat{G}_0 \star \partial_{\pi_{i}}\hat{Q}_0  \right)^<  +{\rm c.c.}\Big)=0
\end{eqnarray}
Notice that the property proven here is not the complete topological invariance unlike the case of equilibrium at $T=0$. The value of $\cal N$ may be changed smoothly under the change of distribution function $f(\pi_0)$. Moreover, its value is modified also when $f(\pi_0)$ remains unchanged but the terms in $\cal N$ proportional to the derivative of $f$ gives valuable contribution.

\section{Hall conductivity for the system of massive $2$D Dirac fermions}
\label{SectI}

The system of massive $2$D Dirac fermions in equilibrium at zero temperature  has been concerned in Sect. \ref{Sect2DDirac}.  It corresponds to
$$
	Q = {\bm 1} \omega -v_F (\sigma^1 \pi_1 + \sigma^2 \pi_2 + \sigma^3 m).
$$
Here $m$ is a mass-type parameter, $\si_i$ are Pauli matrices, and ${\bm 1}$ is a unit $2\times 2$ matrix. In equilibrium at $T=0$ Hall conductivity is given by
$$
	\sigma_H = \rv{-}\frac{{\cal N}}{2\pi}.
$$
with \cite{Volovik2003}
$$
	{\cal N}^{(0)} = \frac{1}{2} {\rm sign} \, m.
$$
Recall that for purely two-dimensional systems these fermions always come in pairs, and the total value of $\cal N$ is integer rather than half-integer.

Now let us calculate corrections to $\sigma_H$ at finite temperatures using the developed formalism. For simplicity we consider the case when Fermi energy is set to zero.
Our starting point is Eq. (\ref{C14}) for Hall conductivity.
The Hamiltonian can be written as follows:
$$
H = v_F \left(\begin{array}{cc}m &  p_1 - \ii p_2 \\  p_1 +\ii p_2 & - m \end{array}\right)
$$
In the following for simplicity we will consider the case $v_F = 1$ (the nontrivial value of Fermi velocity may be easily restored in the final answer). The eigenvalues of this Hamiltonian are $ E_\pm({p}) =\pm \sqrt{|{p}|^2+m^2}$. The corresponding eigenvectors are
$$
|n \rangle \equiv
	 |a {p}\rangle =  \frac{|{p}|}{\sqrt{2\sqrt{|{ p}|^2+m^2} (\sqrt{|{ p}|^2+m^2}+ a m)}}\left(\begin{array}{c} 1  \\ - \frac{m -a \sqrt{|{ p}|^2+m^2}}{p_x +\ii p_y} \end{array}\right)|{ p}\rangle
$$
Here $a=  \pm 1 $, with $+1$ corresponding to the conductance band with positive energy while $-1$ marks valence band with negative energy. Momenta eigenstates are normalized to $1$ in discrete space, $\langle {q}|{p}\rangle = (2\pi)^2 \delta^{(2)}({q}-{p})/{\cal V}$.

Since velocity operator $\hat{v}_k = \sigma_k$ does not contain momentum, its matrix elements between the states with definite momenta $p$ and $q$ contain a delta-function $\delta^{(2)}({p-q})$. In Eq. (\ref{C14}) each of two sums over the quantum states is to be substituted by ${\cal V}\sum_{a \in\{c,v\}} \int \frac{d^2 p}{(2\pi)^2}$.  We also denote
$$
\langle a,{p}|\hat{v}_i|b,{ p}^\prime\rangle \langle b,{ p}^\prime|\hat{v}_j|a,{p}\rangle
	= \langle a,b;i,j;{p} \rangle \frac{(2\pi)^2}{\cal V} \delta^{(2)}({p}-{p}^\prime) \frac{(2\pi)^2}{\cal V} \delta^{(2)}({p}-{p}^\prime)
	= \langle a,b;i,j;{p} \rangle \frac{(2\pi)^2}{\cal V} \delta^{(2)}({p}'-{p}).
$$
In the last equality we used the fact that the factor $\frac{(2\pi)^2}{\cal V} \delta^{(2)}(0)$ is to be replaced by unity, which becomes clear if we consider the system inside a large but finite rectangular box with periodic boundary conditions and replace the integral over momentum by the sum over its discrete values.  Furthermore, it is easy to see that $ \langle a,b;i,j; {p}\rangle =\langle a,b;j,i; {p}\rangle$. Using these notations we may rewrite Eq. (\ref{C14}) as follows
\begin{eqnarray}
\sigma_H= \rv{+} \frac{\ii}{4}\int \frac{d^2p}{(2\pi)^2} \sum_{a,b = \pm 1} \Big[ \frac{f(E_a({p}))-f(E_b({p}))}{(E_a({p})-E_b({p}))^2}  \Big]
     \epsilon_{ij} \langle a,b;i,j;{p}\rangle +c.c.,
\end{eqnarray}
where $f$ is Fermi distribution.

Using the given above explicit expressions for the 2D Dirac spinors, we obtain
$\epsilon_{ij} \langle +,-;i,j;{p}\rangle =2\ii{\rm Im} \langle +,-;1,2;{p}\rangle =2\ii m/\sqrt{{p}^2+m^2}$.
We represent $\sigma_H = I/(8\pi^2)$.  Here
\begin{eqnarray}
I &=& -\ii\int d^2p \sum_{a,b}  \frac{f(E_a({p}))-f(E_b({p}))}{(E_a({p})-E_b({p}))^2}
     \epsilon_{ij} \langle a,b;i,j; {p}\rangle    \nonumber\\
   &=& -2\ii\int d^2p \frac{f(E_+({p}))-f(E_-({p}))}{(E_+({p})-E_-({p}))^2} \epsilon_{ij} \langle c,v;i,j; {p}\rangle \nonumber\\
   &=& -2\ii\int d^2p \Big(\frac{1}{e^{-\beta\sqrt{{p}^2+m^2}}+1}-\frac{1}{e^{\beta\sqrt{{p}^2+m^2}}+1} \Big)
                   \frac{1}{4({p}^2+m^2)}\frac{2\ii m}{\sqrt{{p}^2+m^2}}   \nonumber\\
   &=&  \int d^2p \frac{m}{({p}^2+m^2)^{3/2}}\, {\rm th}\(\frac{\beta\sqrt{{p}^2+m^2}}{2} \)
\end{eqnarray}
Then, after changing variables we obtain the following expression for Hall conductivity:
\begin{eqnarray}
\sigma_H &=&\rv{-} \frac{\alpha}{4\pi} \int_{|\alpha|}^{+\infty} \frac{du}{u^2 \, {\rm th}(u/2)}.
\end{eqnarray}
where $\alpha=v_F\beta m\equiv  v_F m/T$ is dimensionless (we restored here Fermi velocity). This expression tends to $\rv{-}\frac{1}{4\pi}{\rm sign}\,m $ at $T\to 0$.

\end{document}